\documentclass[twocolumn,letterpaper]{aa}
\usepackage[dvips]{graphicx}
\usepackage{parskip}
\usepackage{rotating}
\usepackage{txfonts}

\newcommand{\mic}{$\mu$m}
\newcommand{\mum}{$\mu$m}

\begin{document}

\title{Mid-IR observations of circumstellar disks\thanks
      {Based on observations made at the European Southern Obser\-vatory, 
       obtained under program IDs 076.C-0634(A) and 077.C-0054(A), 
       during TIMMI2 technical time as well as collected from the ESO/ST-ECF 
       Science Archive Facility}}

\subtitle{Part III: A mixed sample of PMS stars and Vega-type objects}

\titlerunning{Mid-IR observations of circumstellar disks (Part III)}

\author{O.~Sch\"utz
        \inst{1}
        \and
        G.~Meeus
        \inst{2}
        \and
        M.~F.~Sterzik
        \inst{1}
        \and
        E.~Peeters
        \inst{3, 4, 5}
        }

\offprints{oschuetz@eso.org}

\institute{European Southern Observatory, Alonso de Cordova 3107,
           Santiago 19, Chile
           \and
           Astrophysikalisches Institut Potsdam, An der Sternwarte~16,
           D-14482 Potsdam, Germany
           \and
           NASA Ames Research Center, MS 245-6, Moffett Field, CA 94035,
           USA
           \and
           SETI Institute, 515 N. Whisman Road, Mountain View, CA 94043, 
           USA
           \and
           The University of Western Ontario, London, ON N6A 3K7, Canada
           }

\date{Received / Accepted}

\abstract{

We present new mid-infrared spectra for a sample of 15 targets (1 
FU~Orionis object, 4 Herbig~Ae stars, 5 T~Tauri stars and 5 Vega-type 
stars), obtained with the TIMMI2 camera at La~Silla Observatory (ESO). 
Three targets are members of the $\beta$~Pic moving group 
(\mbox{\object{HD 155555}}, \mbox{\object{HD 181296}} and \object{HD 319139}). 
PAH bands are observed towards the T~Tauri star \object{HD 34700} and the 
Herbig~Ae star \object{PDS 144 N}. For \object{HD 34700}, the band profiles 
indicate processed PAHs. The spectrum of the Vega-type object $\eta$~Corvi 
(\object{HD 109085}), for which a resolved disk at sub-mm wavelengths is 
known, is entirely stellar between 8--13~$\mu$m. Similarly, 
no indication for circumstellar matter at mid-infrared wavelengths is 
found towards the Vega-like stars \mbox{\object{HD 3003}}, 
\mbox{\object{HD 80951}}, \object{HD 181296} and, surprisingly, the T~Tauri 
system \mbox{\object{HD 155555}}. 

The silicate emission features of the remaining 
eight sources are modelled with a mixture of silicates of different 
grain sizes and composition. Unprocessed dust dominates \object{FU Ori}, 
\object{HD 143006} and \object{CD-43 344}. Large amorphous grains are the 
main dust component around \object{HD 190073}, \object{\mbox{HD 319139}},
\object{KK Oph} and \object{PDS 144 S}. Both small grains and crystalline
dust is found for the Vega-type \object{HD 123356}, with a dominance of small 
amorphous grains. We show that the infrared emission of the binary 
\object{HD~123356} 
is dominated by its late-type secondary, but optical spectroscopy is still
required to confirm the age of the system and the spectral class of the 
companion. For most targets this is their first 
mid-infrared spectroscopic observation. We investigate trends between 
stellar, disk and silicate properties and confirm correlations of previous 
studies. Several objects present an exciting potential for follow-up 
high-resolution disk studies.


          }

\maketitle

\section{Introduction}
\label{sect:introduction}

\begin{sidewaystable*}
  \begin{center}
  \caption{Stellar parameters and known fluxes of our target sample.}
  \linespread{1.3}
  \selectfont
  \bigskip
  \setlength\tabcolsep{6.1pt}
  {\begin{tabular}{lccccrcrcrcccrr}
    \hline
    \hline
    Object     &  Class      &  Spectral Type &  T$_{\mathrm{adopted}}$
               &  (log g)$_{\mathrm{adopted}}$&  V
               &  F$_{\mathrm{12 \, \mu m}}$  
               &                Age           &  Ref.  & $d$       &  Ref.  
               &  $A_v$      &  Ref.          &  Sep.              &  Ref. \\

               &             &                &  [K]
               &                              &  [mag]
               &  [Jy]                        
               &                [Myr]         &        &  [pc]     &       
               &  [mag]      &                &  [$''$]            &       \\
    \noalign{\smallskip}
    \hline
    \noalign{\smallskip}
    HD 3003    &  Vega       &  A0V           &  11000
               &  4.5                         &  5.07
               &  0.48                        
               &       50 $_{- ...}^{+ 197}$  &  (1)   &  47       &  (9)  
               &  ...        &                &  0.1 (?)           &  (26) \\

    HD 34700 A &  TTS        &  G0IVe         &  6000
               &  4.5                         &  9.15
               &  0.60                        
               &                ...           &        &  ...      &  (10)  
               &  0.68       &  (19)          & SB\,/\,5.2\,/\,9.2 &  (27) \\

    HD 80951 A &  Vega       &  A1V           &  9500
               &  4.5                         &  6.00
               &  0.44
               &  ...                         &        &  220      &  (9)
               &  ...        &                &0.3\,/\,7.1\,/\,48.2&  (26) \\

    HD 109085  &  Vega       &  F2V           &  7000
               &  4.5                         &  4.31
               &  2.18                        
               &      1000 $_{- 300}^{+ 300}$ &  (2)   &  18       &  (11)   
               &  0.12       &  (20)          &  ...               &       \\

    HD 123356 AB &  Vega     &  G1V\,/\,M?    &  6000 / 4000
               &  4.5                         &  10.0 / 12.2
               &  1.36                      
               &                ...           &        &  ...      &  (12)
               &  0.16       &  (20)          &  2.5               &  (28) \\

    HD 143006  &  TTS        &  G5Ve          &  5750
               &  4.5                         &  10.21
               &  0.86                        
               &                $\sim$1       &  (3)   &  82       &  (13) 
               &  0.62       &  (21)          &  ...               &       \\

    HD 155555 ABC &  TTS     & G5IV\,/\,K0IV-V\,/\,M4.5           &  5750 / ...
               &  4.5                         &  6.88 / 13.0
               &  0.69
               &        12 $_{- 4}^{+ 8}$     &  (4)   &  32       &  (9)
               &  ...        &                &  SB\,/\,33.0       &  (26) \\

    HD 181296 A &  Vega      &  A0Vn          &  9500
               &  4.5                         &  5.03
               &  0.54
               &        12 $_{- 4}^{+ 8}$     &  (4)   &  48       &  (9)
               &  0.12       &  (20)          &  4.1               &  (29) \\

    HD 190073  &  HAeBe      &  A2IIIpe       &  9500
               &  4.5                         &  7.82
               &  7.16                        
               &      1.2 $_{- 0.6}^{+ 0.6}$  &  (5)   & $>$ 290   &  (14) 
               &  0.19       &  (19)          &  ...               &       \\

    HD 319139  &  TTS        &  K5Ve / K7Ve   &  4250
               &  4.5                         &  10.44
               &  0.45
               &       5.5 $_{- 1.5}^{+ 1.5}$ &  (6)   &  83       &  (15)
               &  $\leq$ 0.12&  (22)          &  SB                &  (30) \\

    CD-43 344  &  TTS (?)    &  M2            &  4500
               &  4.5                         &  9.42
               &  3.73                     
               &                ...           &        &  ...      &  (10)   
               &  0.0        &  (20)          &  ...               &       \\

    FU Ori A   &  FUOR       &  G0II          &  ...
               &  4.5                         &  8.94
               &  5.95
               &                $<$0.3        &  (7)   & 450       &  (16)
               &  2.2        &  (23)          &  0.5               &  (31) \\

    KK Oph A   &  HAeBe      & A6Ve           &  8750
               &  4.5                         &  9.4\,--\,12.9 
               &  9.87
               &         7 $_{- 0.5}^{+ 1.0}$ &  (8)   & 160       &  (17)
               &  1.6        &  (24)          &  1.6               &  (32) \\

    PDS 144 N  &  HAeBe      &  A2IV          &  8750
               &  4.5                         &  14.2
               &  ...
               &                ...           &        & $\sim$1000&  (18)
               &  3.0        &  (25)          &  $\sim$5.0         &  (33) \\

    PDS 144 S  &  HAeBe      &  A5V           &  8750
               &  4.5                         &  13.1
               &  ...
               &                ...           &        & $\sim$1000&  (18)
               &  1.1        &  (25)          &  $\sim$5.0         &  (33) \\
  \noalign{\smallskip}
  \hline
  \end{tabular}}
  \end{center}
  \vspace{0.2cm}
           References: {\bf
           Spectral types} are obtained either from SIMBAD or one of the 
           cited papers, if a more accurate classification is available 
           in the literature. {\bf T$_{\mathrm{adopted}}$} and {\bf
           (log~g)$_{\mathrm{adopted}}$} list the temperature and surface 
           gravity which we used for the Kurucz atmosphere models. 
           {\bf V-band} magnitudes and {\bf IRAS} 12~$\mu$m fluxes are 
           taken from the SIMBAD database. KK~Oph shows variability in 
           the visual. PDS~144 is not 
           resolved by IRAS; its V-band fluxes were given by Perrin et 
           al.\ (\cite{Perrin}). {\bf Stellar ages:} (1)~Song et 
           al.\ (\cite{Song01}), (2)~Wyatt et al.\ (\cite{Wyatt}) and
           references therein, (3)~Dunkin et al.\ (\cite{Dunkin}) and
           references therein, (4)~mean age of the $\beta$~Pic moving 
           group as given in Zuckerman \& Song 
           (\cite{Zuckerman01}), (5)~Catala et al.\ (\cite{Catala}), 
           (6)~Quast et al.\ (\cite{Quast}), (7)~e.g.\ D'Angelo et 
           al.\ (\cite{Angelo}), (8)~Carmona et al.\ (\cite{Carmona}). 
           Errors in the age determinations are quite heterogeneous, 
           depending on the cited work. {\bf Distances} are taken from 
           (9)~Hipparcos parallax, (10)~uncertain Hipparcos parallax, 
           (11)~Wyatt et al.\ (\cite{Wyatt}), (12)~The Hipparcos parallax 
           was recently removed from SIMBAD and is likely not firm, 
           (13)~Sylvester et al.\ (\cite{Sylvester96}), (14)~van den 
           Ancker et al.\ (\cite{Ancker}), (15)~Quast et al.\ (\cite{Quast}), 
           (16)~e.g.\ Wang et al.\ (\cite{Wang}), (17)~Hillenbrand et 
           al.\ (\cite{Hillenbrand}), (18)~Perrin et al.\ (\cite{Perrin}) 
           and references therein. {\bf Extinctions:} (19)~van den Ancker 
           et al.\ (\cite{Ancker}), (20)~Sylvester \& Mannings 
           (\cite{Sylvester00}), (21)~Malfait et al.\ (\cite{Malfait98}), 
           (22)~Hutchinson et al.\ (\cite{Hutchinson}), (23)~Kenyon et 
           al.\ (\cite{Kenyon88}), (24)~Carmona et 
           al.\ (\cite{Carmona}), (25)~Perrin et al.\ (\cite{Perrin}).
           {\bf Stellar separations in multiple systems:} (26)~Dommanget
           \& Nys (\cite{Dommanget}), (27)~Sterzik et al.\ (\cite{Sterzik}),
           (28)~Worley \& Douglass (\cite{Worley}), (29)~Lowrance et
           al.\ (\cite{Lowrance}), (30)~Quast et al.\ (\cite{Quast}),
           (31)~Wang et al.\ (\cite{Wang}), (32)~Leinert et 
           al.\ (\cite{Leinert97}), (33)~Perrin et al.\ (\cite{Perrin}).
           'SB' denotes a spectroscopic binary. {\bf Note on the multiplicity
           of HD~155555:} Although the SB components are not spatially 
           resolved, we label them as A and B, since this naming convention
           existed already in the literature.
  \linespread{1}
  \selectfont
  \label{table:targets}
\end{sidewaystable*}

Up to now a full understanding of dust evolution in pre-main sequence 
(PMS) disks and the survival of debris disks is lacking. Both disk 
types need to be distinguished in terms of their dust evolution. For an 
in-depth discussion of gas and dust evolution in circumstellar disks, we 
refer to the review by Meyer et al.\ (\cite{Meyer}).

Primordial PMS 
disks largely reflect the grain size distribution of the interstellar 
medium (ISM), while more evolved PMS disks experience grain growth and 
crystallisation. However, there is no global 
correlation of dust properties with age, not even when considering individual 
spectral types. Schegerer et al.\ (\cite{Schegerer}) re-analysed the 
mid-infrared silicate emission features of 27 T~Tauri stars (TTS) and found 
that the overall degree of crystallinity in T~Tauri disks correlates 
neither with stellar age nor luminosity. Through resolved interferometric 
observations, van Boekel et al.\ (\cite{Boekel04}) showed that a higher 
amount of crystalline dust is seen in close proximity to the star than in 
the outer disk regions. In an analysis of 40 TTS and 7 HAeBe stars, 
Kessler-Silacci et al.\ (\cite{Kessler}) found a correlation with spectral 
type for the ratio strength-to-shape in the N-band silicate feature, but no 
relationship with stellar age or disk evolutionary status. Therefore, the 
observed grain sizes in PMS disks must also depend on other factors like, 
e.g., close companions, or turbulence and gas content in the disk, while 
the exact dependencies thereof remain unknown. Kessler-Silacci 
et al.\ (\cite{Kessler07}) demonstrated that the disk region which is 
probed by the 10~$\mu$m silicate emission features, is related to stellar 
luminosity, what could explain the correlation of grain 
size with spectral type.

\begin{table*}[t]
  \begin{center}
  \caption{Results from TIMMI2 photometry and spectroscopy.}
  \setlength\tabcolsep{6.2pt}
  \begin{tabular}{lcccccccr}
    \hline
    \hline
    Object              &  Origin
    &  Airmass (N-spec) &  t$_{\mathrm{int}}$  (N-spec)      & 
                           F$_{\mathrm{N1}}$                 &
                           F$_{\mathrm{N10.4}}$              &
                           F$_{\mathrm{N11.9}}$              & Ref. &
                           N-spec features                   \\
                        
                        & 
    &                   &  [min]                             &      
                           [Jy]                              & 
                           [Jy]                              &
                           [Jy]                              &      &
                                                             \\
    \hline
    HD 3003             &  technical
    &  1.2              &      24          
    &  0.58 $\pm$ 0.03  &  0.32 $\pm$ 0.03  &  0.20 $\pm$ 0.10      & ...
    &  stellar                          \\

    HD 34700            &  technical
    &  1.2\,--\,1.5     &      60         
    &  ...               &  ...               &  0.79 $\pm$ 0.29      & ...
    &  PAH                              \\

    HD 80951            &  077.C-0054
    &  1.7              &      53       
    &  ...               &  ...               &  0.44 $\pm$ 0.03      & (1)
    &  stellar                          \\

    HD 109085           &  archive
    &  1.1              &      12       
    &  ...               &  ...               &  1.51 $\pm$ 0.24      & (2)
    &  stellar                          \\

    HD 123356           &  archive
    &  1.0              &      18       
    &  ...               &  1.51 $\pm$ 0.05  &  1.68 $\pm$ 0.06      & ...
    &  silicate emission                \\

    HD 143006           &  archive
    &  1.1              &      24       
    &  ...               &  ...               &  0.86 $\pm$ 0.06      & (1)
    &  silicate emission                \\

    HD 155555           &  077.C-0054
    &  1.3              &      58       
    &  ...               &  ...               &  0.46 $\pm$ 0.03      & ...
    &  stellar                          \\

    HD 181296           &  077.C-0054
    &  1.2              &      64       
    &  0.59 $\pm$ 0.05  &  ...               &  ...                   & ...
    &  stellar                          \\

    HD 190073           &  technical
    &  1.5              &      12       
    &  ...               &  ...               &  7.16 $\pm$ 0.43      & (1)   
    &  silicate emission                \\       

    HD 319139           &  077.C-0054
    &  1.0              &     160       
    &  ...               &  ...               &  0.36 $\pm$ 0.03      & ...
    &  silicate emission                \\

    CD-43 344           &  technical
    &  1.1              &      24       
    &  3.78 $\pm$ 0.05  &  3.82 $\pm$ 0.05  &  3.04 $\pm$ 0.05      & ...
    &  silicate emission                \\

    FU Ori              &  technical  
    &  1.4              &      20       
    &  ...               &  6.00 $\pm$ 0.03  &  5.61 $\pm$ 0.03      & ...
    &  silicate emission                \\

    KK Oph              &  archive
    &  1.2              &      12       
    & 10.09 $\pm$ 0.05  &  ...               & 10.71 $\pm$ 0.08      & ...
    &  silicate emission                \\

    PDS 144 N           &  077.C-0054
    &  1.0              &      32       
    &  ...               &  ...               &  1.74 $\pm$ 0.09      & (3)
    &  PAH                              \\

    PDS 144 S           &  077.C-0054
    &  1.0              &      13       
    &  ...               &  ...               &  2.49 $\pm$ 0.13      & (3)
    &  silicate emission                \\
    \noalign{\smallskip}
    \hline
  \end{tabular}
  \end{center}
  \vspace{0.1cm}
           Results from TIMMI2 photometry and spectroscopy are merged in
           this table. The airmass and TIMMI2 integration times refer to 
           N-band spectroscopy. Mid-IR photometry was obtained in the 
           specified TIMMI2 filters with central wavelengths 8.6~$\mu$m 
           (N1), 10.3~$\mu$m (N10.4) and 11.6~$\mu$m (N11.9). Errors 
           represent the accuracy of the aperture photometry. 
           Where no own measurements were available, e.g, for some archive
           data, we adapted data from (1)~IRAS, (2)~Wyatt et 
           al.\ (\cite{Wyatt}) or (3)~Perrin et al.\ (\cite{Perrin}).
  \label{table:obs-results}
  \bigskip
  \medskip
\end{table*}

Debris disks -- also referred to as Vega-type objects -- are often 
characterised by the lack of gas, so that the dust dynamics is no
longer influenced by that of the gas. The presence of small dust particles 
in such disks points towards a continuous replenishment of the dust grains, 
e.g.\ by collisions between planetesimals, asteroids or comets. 
Indeed, through collisional cascades significant amounts of small grains can 
be produced, down to grain sizes which are easily removed by radiation 
pressure on short timescales, and therefore, without replenishment would not
be expected to be present in the disk. Song et 
al.\ (\cite{Song05}) found a surprisingly strong silicate feature for the 
$\sim$300~Myr old star \mbox{\object{BD+20 307}}, which they explain by the 
release of secondary dust after the destruction of asteroids. The emission 
peak shortward of 10$~\mu$m in its spectrum indicates rather small grains. 
In Sch\"utz et al.\ (\cite{paper2}, hereafter Paper\,II), we discussed a 
similar - but much younger - main-sequence object, \object{HD 113766}, which 
shows large amounts of crystalline secondary dust, possibly released by 
collisions in a planetesimal belt. By applying spatially 
resolved spectroscopy to \object{$\beta$~Pic}, Okamoto et 
al.\ (\cite{Okamoto}) were able to trace the origin of secondary dust towards 
the location of putative planetesimal belts. \mbox{\object{HD 69830}} is 
another system with secondary debris dust in form of warm, small, crystalline 
silicates (Beichman et al.\ \cite{Beichman}), possibly resulting from the 
disruption of an asteroid within 1~AU from the star (Lisse et 
al.\ \cite{Lisse}). Interestingly, this system is also known to host 
three Neptune-sized planets (Lovis et al.\ \cite{Lovis}). 

The present paper continues our previous mid-infrared (MIR) spectroscopic 
studies 
of circumstellar disks. In Sch\"utz et al.\ (\cite{paper1}, hereafter 
Paper\,I) we described our strategy and dust emission models, and analysed 
a sample of pre-main sequence stars. Vega-type objects with debris disks 
were studied with the same methods in Sch\"utz et al.\ (\cite{paper2}, 
Paper\,II). 

\begin{figure}[t]
  \centering
  \includegraphics[scale=0.6, angle=90]{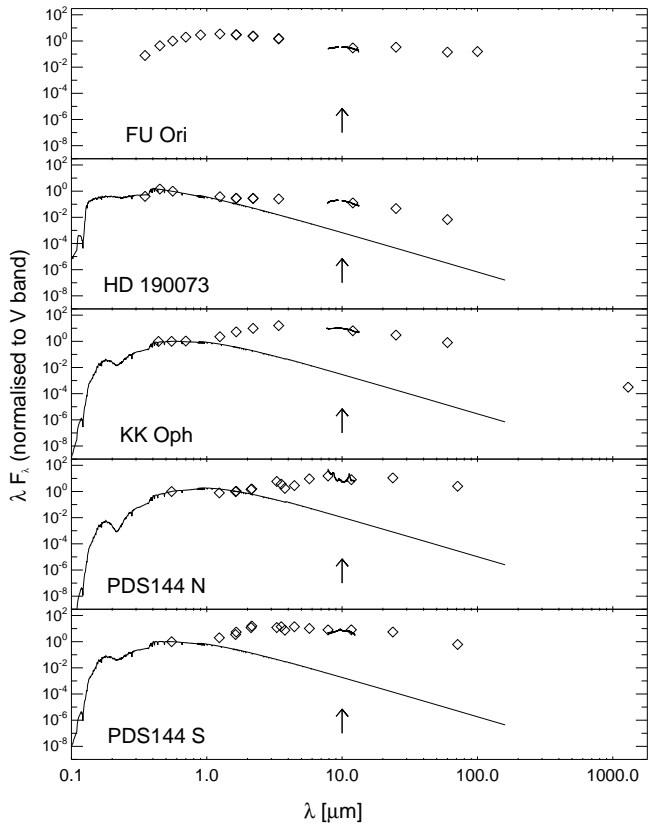}
  \caption{Spectral energy distributions for FU Ori and the Herbig~Ae 
           stars. A Kurucz model is overplotted for all targets to 
           represent the stellar contribution to the SED, apart from 
           FU~Ori as justified in the text. The profile of the N-band 
           spectrum is indicated in the SEDs (at the position of the 
           arrow) to show their agreement 
           with the photometry. For PDS 144 N and S all IR fluxes are 
           taken from Perrin et al.\ (\cite{Perrin}).}
  \label{fig:sed_herbig}
  \bigskip
  \medskip
\end{figure}

This contribution describes new mid-infrared spectra of circumstellar 
disks. We selected targets with infrared excess that have not yet been 
discussed in depth in the literature, i.e.\ objects which were 
suspected to be disk candidates, but for which no MIR spectroscopy had
been obtained before. Several isolated targets were unobserved by Spitzer, 
as they were not included 
in the surveys of open clusters and moving groups. We re-observed 
targets where newer or higher SNR data were required, as for 
\object{\mbox{HD 34700}}. A MIR 
spectrum of \object{\mbox{KK Oph}} had already been published, however, it 
was not analysed with the methods we present here. We included the 
well studied object \object{\mbox{FU Ori}} to enable a comparison with 
other FU Orionis targets from Paper\,I. Table~\ref{table:obs-results}
summarises our targets and their stellar parameters. In 
Sect.\,\ref{sect:observation} 
we describe the observations and data reduction. Sect.\,\ref{sect:results} 
discusses the spectral energy distributions (SEDs) and methods 
for analysing the emission feature in the 8--13~$\mu$m spectra. 
In Sect.\,\ref{sect:discussion} we describe observed trends in the dust
properties. A summary of our results is given in Sect.\,\ref{sect:conclusions}.
The known properties of each target are described in detail in the appendix. 
It is our goal to contribute a piece to the puzzle of disk and dust 
evolution around young stars.

\begin{figure}[t]
  \centering
  \includegraphics[scale=0.6, angle=90]{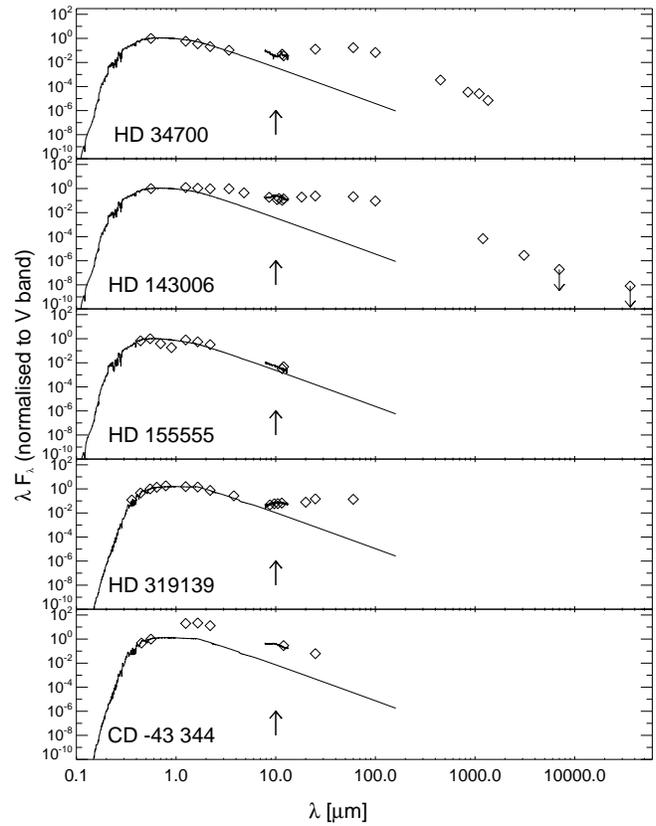}
  \caption{The spectral energy distributions for the T~Tauri stars. Remarkably,
  \object{HD 155555} shows no excess emission in the infrared.}
  \label{fig:sed_ttauri}
  \bigskip
  \bigskip
\end{figure}

\section{Observations and data reduction}
\label{sect:observation}

{\bf TIMMI2:}
The principal data described in this work were obtained with the ESO TIMMI2 
camera (K\"aufl et al.\ \cite{Kaufl}) at La~Silla Observatory. We performed
most of the observations during run 077.C-0054(A) in June 
2006, supported by further data retrieved from the ESO/ST-ECF Science 
Archive\footnote{http://archive.eso.org}, which originally had been 
observed in March 2003 (targets \mbox{\object{HD 109085}}, \object{HD 123356}
and \object{\mbox{HD 143006}}; ESO program 070.C-0743(A); PI J.-C.~Augereau) 
and July 2003 (target \mbox{\object{KK Oph}}; ESO program 071.C-0438(A); PI 
H.~Linz). Further data were collected during TIMMI2 technical nights in 
October 2005 and January 2006. Table~\ref{table:obs-results} summarises the 
origin of all data. 

Imaging data were obtained with a 0.20$''$/pixel scale, corresponding to
a 64$''$\,x\,48$''$ field of view. For most targets the nodding direction
was perpendicular to the chop throw, resulting in four images of the source
in the co-added frame, each carrying 25\% of the flux. Some archive data, 
however, had the nod parallel to the chop, giving three images of the 
source, with the central one carrying 50\% of the flux.
We applied aperture photometry with radii $r_{\mathrm{ap}}$ between 2.6 
pixel and 4.7 pixel, depending on the target. The 
optimal radii were derived from curves of growth and averaged over the radii 
obtained for each of the beams, to account for detector 
inhomogeneities. As error in the photometry we give the mean of the 
measurement errors for each beam. The background level was determined 
by measuring the 
mean sky value per pixel around each beam in an annulus centered on the 
star with inner radius $r_{\mathrm{ap}} + 10$~pixel and outer radius 
$r_{\mathrm{ap}} + 20$~pixel (resulting in a maximum outer radius of 
5$''$, while the chopping amplitude is 10$''$). Photometric standard stars
were selected from a list of MIR standards by Cohen (\cite{Cohen}) and 
observed approximately each 2 hours.

N-band spectra were obtained between 8--13~$\mu$m applying a standard 
chopping and nodding technique along the 1.2$''$ slit (for archive data 
and technical time) or the 3$''$ slit (run 077.C-0054(A)) with a throw of 
10$''$. The spectral resolution of the used 10~$\mu$m low-resolution grism 
is 0.02 $\mu$m\,/\,pixel ($\lambda$\,/\,$\delta \lambda \sim 160$). Airmass 
and on-source integration times of the observations are listed in 
Table~\ref{table:obs-results}. Standard stars for telluric correction 
and flux calibration were selected from the list of Cohen (\cite{Cohen}) 
and observed close in time and airmass. 

The spectroscopic frames were de-biased and checked for a correct slit 
alignment along the detector, i.e.\ whether the spectrum is aligned with
the x-axis. After co-adding the positive and negative 
nodding positions, the extraction profile is determined from the cross 
dispersion profile: the target spectrum is extracted along a profile width 
where the SNR exceeds 2.5 times the noise of the background. 
We furthermore applied an airmass correction described in Sch\"utz \& 
Sterzik (\cite{Schuetz}) and correlate the spectrophotometric calibration 
with the obtained N-band photometry. The wavelength calibration is 
optimised using the atmospheric absorption features of ozone 
(9.58~$\mu$m) and CO$_2$ (11.73~$\mu$m, 12.55~$\mu$m) in addition to the 
given TIMMI2 wavelength calibration table, i.e.\ shifts were applied to  
minimise residual division between the atmospheric features in standard
stars and target spectra. When atmospheric conditions 
were unstable between observations of target and standard star, leaving 
uncorrectable ozone features around 9.5~$\mu$m, we cut these remnant 
features from the spectra. By comparing data from different nights and 
applying several standard stars for comparison, the exact position of the 
ozone remnant feature can be identified. Depending on the atmospheric 
quality, this leaves gaps of different sizes up to a maximum range 
between 9.4 and 9.9~$\mu$m. The SNR in our target spectra, averaged
between 8--12~$\mu$m, ranges between 10 and about 70 for all objects,
with a minimum SNR in the spectra of \object{HD 3003} and 
\mbox{\object{HD 319139}}, while the highest SNR is obtained for 
\object{FU Ori} and \mbox{\object{KK Oph}}.

\begin{figure}[t]
  \centering
  \includegraphics[scale=0.60, angle=90]{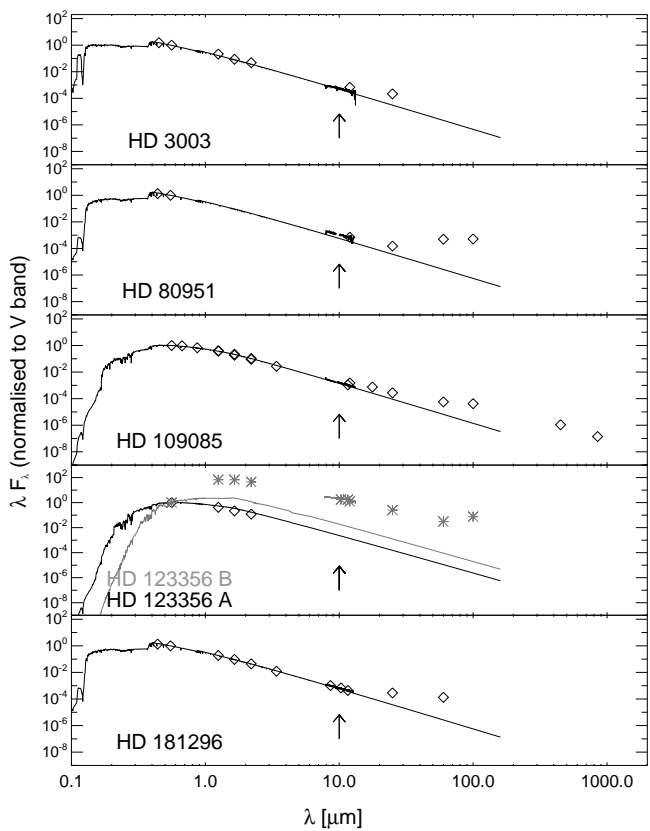}
  \caption{Spectral energy distributions of the Vega-type stars. For 
           the binary HD~123356 the SED of the primary ({\sl black curve}) is
           entirely stellar, while the excess emission is dominated 
           by the secondary ({\sl grey curve and stars}).}
  \label{fig:sed_vega}
  \bigskip
\end{figure}

{\bf Spitzer:}
\object{HD 34700} is also observed with the IRS on board the Spitzer Space
Telescope (Werner et al.\ \cite{Werner}) (AOR key = 3584768, GTO program 
of J.\ Houck). The spectra were taken with the Short-Low (SL, 5--15~$\mu$m) 
and Short-High (SH, 10--20~$\mu$m) modules (Houck et al.\ \cite{Houck}) 
with a resolving power of respectively $\mathrm R = $~60--128 and 600. The 
basic data reduction was performed by the standard IRS pipeline S13 at the 
Spitzer Science Center (SSC). Subsequently, bad pixels were removed with 
IRSCLEAN (available from the SSC 
website\footnote{http://ssc.spitzer.caltech.edu}) and a standard 
point-source extraction has been performed using SMART (Higdon et al.\
\cite{Higdon}): tapered column extraction for SL data (i.e.\ the width of 
the extraction scales with wavelength and equals 7.2$''$ at 6~$\mu$m); full 
aperture extraction for SH data (4.7$''$\,x\,11.3$''$). 

{\bf Further auxiliary data:}
During the discussion of two targets -- \object{HD 109085} and 
\object{HD 123356} -- we will refer to individual data obtained either with 
VISIR (Lagage et al.\ \cite{Lagage}) or SOFI (Moorwood et 
al.\ \cite{Moorwood}).

\section{Analysis}
\label{sect:results}

\subsection{Spectral Energy Distributions}
\label{sect:sed}

Together with fluxes from the literature in the passbands UBVRI, JHK 
(2MASS), 12, 25, 60 and 100~$\mu$m (IRAS) we used the TIMMI2 photometry 
in Table~\ref{table:obs-results} to construct a spectral energy
distribution (SED) for our targets. In 
Figs.\,\ref{fig:sed_herbig}--\ref{fig:sed_vega} the resulting optical 
to mm spectral energy distributions are displayed. To emphasise the 
non-stellar contribution to the SED, we plotted a Kurucz atmosphere 
model (Kurucz \cite{Kurucz}) with the stellar parameters from 
Table~\ref{table:targets}, normalised to the V-band flux. 
For \object{FU Ori} we did not plot 
a Kurucz model, as for this kind of object the V-band flux most likely 
arises from the star plus the hot inner part of the outbursting disk 
(cf.\ Sect.\,\ref{sect:fuor}).
Where required, we reddened the Kurucz model with the extinction 
$\mathrm{A_v}$ given in Table~\ref{table:targets}. The photometry to
construct the SEDs was taken from Hutchinson et al.\ (\cite{Hutchinson}),
Henning et al.\ (\cite{Henning}), Sylvester et al.\ (\cite{Sylvester96}),
Coulson et al.\ (\cite{Coulson}), Malfait et al.\ (\cite{Malfait98}), 
Mannings \& Barlow (\cite{Mannings98}), Sylvester \& 
Mannings (\cite{Sylvester00}), Sylvester et al.\ (\cite{Sylvester01}), 
Jayawardhana et al.\ (\cite{Jayawardhana}), Mamajek et 
al.\ (\cite{Mamajek}), Natta et al.\ (\cite{Natta}), Sheret et 
al.\ (\cite{Sheret}), Wyatt et al.\ (\cite{Wyatt}), Perrin et 
al.\ (\cite{Perrin}), the catalogs JP11, 2MASS, IRAS, and SIMBAD.

The SED of \object{HD 155555} can entirely be explained by stellar 
emission, while for \object{HD 3003}, \object{HD 80951}, 
\object{HD 109085} and \object{\mbox{HD 181296}} the IR excess begins 
longward of 25~$\mu$m. We, therefore, do not consider those objects in 
our dust analysis. For the other targets a near-infrared (NIR) excess 
is seen, starting longward of the L-band for \object{HD 34700} as well 
as \object{\mbox{HD 319139}}, and around 1~$\mu$m for the remaining eight
targets.

\begin{table*}[t]
  \centering
  \caption{Relative mass fractions in percent derived from the model fits 
           to the N-band spectra. ``Amorphous silicates'' indicate the sum of 
           amorphous olivines and pyroxenes. Targets which do not show 
           a silicate feature are not listed.
           }
  \setlength\tabcolsep{14.2pt}
  \begin{tabular}{lcccccccc}
    \hline
    \hline
    Object     &\multicolumn{2}{c}{Amorphous silicate}
               &\multicolumn{2}{c}{Forsterite}  
               &\multicolumn{2}{c}{Silica}& \multicolumn{2}{c}{Enstatite}\\
               &0.1~\mic&2.0~\mic& 0.1~\mic
               &2.0~\mic&0.1~\mic&2.0~\mic&0.1~\mic&2.0~\mic          \\
    \hline
    \noalign{\smallskip}
    HD~123356  &  76.0$_{- 1.3}^{+  1.3}$  &   0.0$_{- 0.0}^{+  0.2}$  
               &   7.0$_{- 0.5}^{+  0.5}$  &   ...  
               &  11.4$_{- 0.3}^{+  0.3}$  &   ...
               &   4.6$_{- 0.6}^{+  0.7}$  &   1.0$_{- 1.0}^{+  1.0}$ 
    \medskip \\
    HD~143006  &  76.4$_{- 4.0}^{+  5.7}$  &   ...  
               &   3.0$_{- 0.2}^{+  0.1}$  &   1.9$_{- 0.6}^{+  0.3}$  
               &   ...                      &   9.5$_{- 0.7}^{+  0.3}$  
               &   1.0$_{- 0.2}^{+  0.3}$  &   8.2$_{- 1.3}^{+  0.4}$ 
    \medskip \\
    HD~190073  &   3.7$_{- 0.6}^{+  0.5}$  &  85.3$_{- 0.2}^{+  0.3}$  
               &   ...                      &   ...
               &   ...                      &   7.2$_{- 0.1}^{+  0.1}$  
               &   3.9$_{- 0.1}^{+  0.1}$  &   ...                     
    \medskip \\
    HD 319139  &  25.7$_{- 17.0}^{+  18.4}$&  63.6$_{- 6.8}^{+  6.5}$  
               &   1.4$_{- 0.4}^{+  0.3}$  &   4.2$_{- 1.7}^{+  1.6}$  
               &   1.3$_{- 1.1}^{+  1.2}$  &   3.8$_{- 1.0}^{+  1.1}$  
               &   ...                      &   ...                     
    \medskip \\
    CD-43 344  &  74.3$_{- 1.4}^{+  0.7}$  &  24.8$_{- 0.6}^{+  0.9}$  
               &   ...                      &   0.9$_{- 0.3}^{+  0.7}$  
               &   ...                      &   ...                    
               &   ...                      &   ...                     
    \medskip \\
    FU Ori     &  87.9$_{- 6.8}^{+  5.6}$  &   ...
               &   0.3$_{- 0.0}^{+  0.2}$  &   9.0$_{- 0.1}^{+  0.7}$  
               &   1.4$_{- 0.1}^{+  0.2}$  &   ...
               &   1.4$_{- 0.1}^{+  0.2}$  &   ...                     
    \medskip \\
    KK Oph     &   9.7$_{- 0.8}^{+  0.8}$  &  70.3$_{- 9.2}^{+  9.3}$  
               &   2.1$_{- 0.1}^{+  0.1}$  &   6.1$_{- 0.4}^{+  0.4}$  
               &   ...                      &  11.8$_{- 0.2}^{+  0.2}$   
               &   ...                      &   ...                     
    \medskip \\
    PDS 144 S  &  30.7$_{- 0.3}^{+  1.4}$  &  54.6$_{- 1.8}^{+  0.4}$  
               &   0.5$_{- 0.1}^{+  0.2}$  &   ...
               &   ...                      &  14.2$_{- 0.1}^{+  0.3}$   
               &   ...                      &   ...                     
    \smallskip \\
    \hline
  \end{tabular}
  \label{tab:masses}
  \bigskip
  \medskip
\end{table*}

For \object{HD 109085} (alias \object{$\eta$ Crv}) we obtained Q1 band 
(17.7~$\mu$m) imaging with VISIR during run 
076.C-0634(A) in March 2006. \object{$\eta$ Crv} was observed using a 
standard chopping technique with a throw of 10$\arcsec$ at an airmass of 
1.03, while the standard star (\object{HD 99167}) was obtained at airmass 
1.35. Standard data reduction and photometric extraction was applied, 
including a correction for the atmospheric extinction. The derived 
photometry is 1.59~Jy ($\pm$~0.10~Jy). Subtraction of the PSF showed that 
the source is not resolved at 17.7~$\mu$m: the FWHM is 6.7 pixel 
(0.50$\arcsec$) for the standard star while only 6.8 pixel (0.51$\arcsec$) for 
\mbox{\object{$\eta$ Crv}}.

We determined from which component of \object{HD 123356} the excess
emission arises. The 2MASS photometry gives the combined fluxes of 
both sources but interestingly, the corresponding 2MASS coordinates seem 
to be centered towards the secondary. With SOFI images obtained during 
thick clouds in May 2006
we could separate the emission and find that the secondary dominates 
the IR. The relative NIR brightness difference between secondary and 
primary amounts to 3.3\,mag in filter NB1.19, 4.0\,mag in NB1.71 and 
FeII-H, and 4.2\,mag in NB2.09. With this information it was possible 
to split the 2MASS fluxes between both components and to plot a resolved 
SED (see Fig.\,\ref{fig:sed_vega}). From the VJHK colors of 
both stars, the primary appears as a F7-G1, in agreement with the 
SIMBAD classification, while the secondary seems to be either a very late 
M-type or it suffers extinction. In TIMMI2 MIR data we only see one 
isolated source. Although astrometry is not possible on TIMMI2, with the 
observations discussed above, we can safely assume that we only see the 
secondary in the MIR. In the SED we plot Kurucz models for both components.

\subsection{Dust properties}
\label{sect:n-band}

\begin{figure}[t]
  \centering
  \includegraphics[scale=0.46, angle=0]{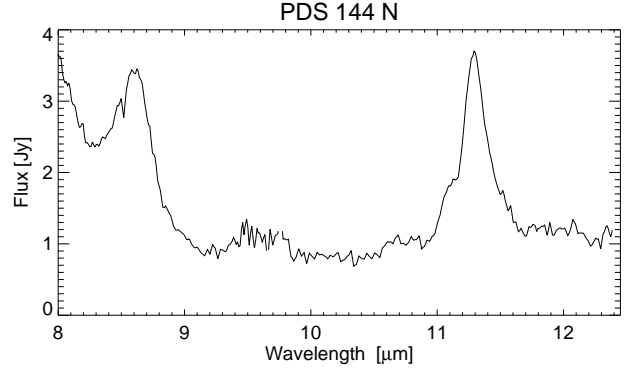}
  \caption{The TIMMI2 spectrum of PDS 144 N, showing the 8.6, 11.0 and 
           11.2~$\mu$m PAH bands.}
  \label{fig:pds144n_1}
  \bigskip
  \medskip
\end{figure}

In Figs.\,\ref{fig:decompo_herbig}--\ref{fig:pds144n_1} 
and~\ref{fig:t2-vs-spitzer}, we show the N-band spectra for our targets. 
To quantify the differences between the sources and to determine 
the composition of the circumstellar dust, we adopt the same procedure 
as described 
in Sect.\,4 of Paper\,I, which is summarised below in 
Sect.\,\ref{sect:silicates}. For further details we refer to Paper\,I.
Since two objects show strong PAH emission, a description of PAH classes
is given in Sect.\,\ref{sect:pahs}.

\subsubsection{Silicates}
\label{sect:silicates}

When stars form out of their parental cloud, the material in
their disks is assumed to have a similar {\it initial} composition as in the
interstellar medium, in which amorphous silicates are the main
component observed at 10~$\mu$m (e.g.\ Kemper et al.\ \cite{Kemper}). 
Changes in the composition, structure and size are expected to occur during 
the subsequent evolution of the star+disk system, possibly leading to a 
planetary system. Laboratory experiments have suggested that, due to
thermal annealing, amorphous silicates gradually turn into crystalline 
forsterite and silica (e.g.\ Rietmeijer \cite{Rietmeijer}; Hallenbeck
\& Nuth \cite{Hallenbeck}). Bouwman et al.\ (\cite{Bouwman01}) found a 
correlation between the amount of forsterite and silica in the disks
of Herbig Ae/Be (HAeBe) stars, indicating that thermal annealing indeed
takes place in these objects. A similar correlation was also found for
the lower-mass T~Tauri stars (e.g.\ Meeus et al.\ \cite{Meeus03}). These
authors also concluded that the dust around HAeBe stars and T~Tauri stars
has very similar characteristics. In Paper\,II, we successfully modelled 
a sample of Vega-type stars with the same dust species as those found 
in young stellar objects. The above-mentioned dust species 
emit in the N band, making this an excellent window to study dust 
evolution in the inner parts of the circumstellar disk.

\begin{figure}[t]
  \centering
  \includegraphics[scale=1.04, angle=90]{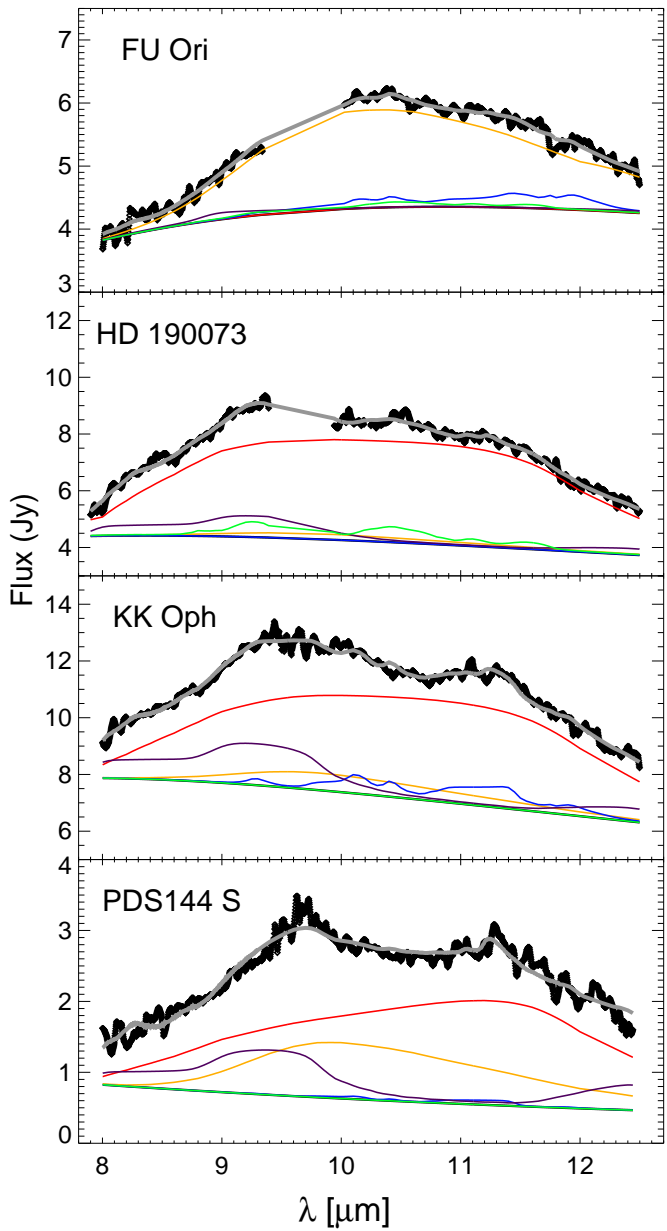}
  \caption{Dust decomposition for FU Ori and the Herbig Ae stars.
           The different linestyles -- or colours in the online 
           version -- represent small amorphous silicates
           ({\sl dash-triple-dots, orange}), large amorphous silicates 
           ({\sl dashed, red}), both small and large grains of crystalline 
           forsterite ({\sl dash-dots, blue}), silica (SiO$_2$, 
           {\sl dotted, purple}) and enstatite ({\sl long curved dashes,
           green}). The very thick black curve corresponds to the observed 
           spectrum. The sum of the different model components is given by 
           the overplotted grey curve.}
  \label{fig:decompo_herbig}
  \medskip
\end{figure}

Our analysis is based on a method that is commonly used when analysing the 
dust features around 10~$\mu$m: fitting a continuum to the spectral range 
outside the feature, and then decomposing the continuum-subtracted emission 
feature by fitting 
a linear combination of absorption coefficients of different dust species, 
for a few selected sizes. For a deeper discussion on this method, we refer 
to Bouwman et al. (\cite{Bouwman01}) and van Boekel et al. (\cite{Boekel05}).
To derive the composition of the circumstellar dust, we model the spectra  
in Figs.\,\ref{fig:decompo_herbig}--\ref{fig:decompo_ttauri} with a linear 
combination of emission features from the following dust species, which 
have been identified in the disks of pre-main sequence stars:

\begin{figure}[t]
  \centering
  \includegraphics[scale=1.04, angle=90]{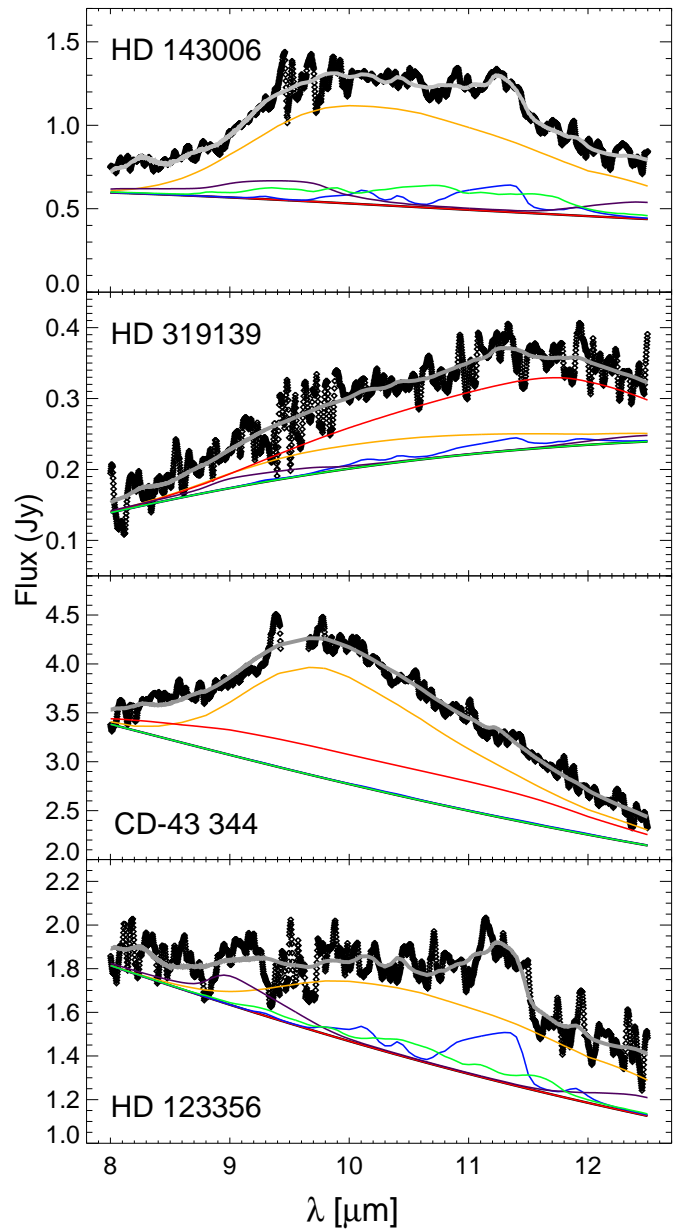}
  \caption{Dust decomposition for the T~Tauri stars (first three panels). 
           See Fig.\,\ref{fig:decompo_herbig} for an explanation of the
           linestyles. The bottom panel shows HD~123356, the only 
           Vega-type object in our sample that shows a silicate feature.}
  \label{fig:decompo_ttauri}
  \bigskip
  \medskip
\end{figure}

\begin{itemize}

\item{
Amorphous olivine ([Mg,Fe]$_2$SiO$_4$) with grain sizes of 0.1 and
2.0~$\mu$m, to which we will refer to as ``small'' and ``large'' grains.
We used absorption coefficients of spherical grains based on laboratory
experiments by Dorschner et al.\ (\cite{Dorschner}).
}

\item{
Amorphous pyroxene ([Mg,Fe]SiO$_3$), as well with grain sizes of 0.1 
and 2.0~$\mu$m. These absorption coefficients are also based on Dorschner 
et al.\ (\cite{Dorschner}).
}

\item{
Crystalline silicates: magnesium forsterite (Mg$_2$SiO$_4$) and enstatite 
(MgSiO$_{3}$). The latter one, however, is a rare find even in young 
sources (cf.\ Paper\,I). Coefficients for forsterite are
from Servoin \& Piriou (\cite{Servoin}), for enstatite from Jaeger et
al.\ (\cite{Jaeger}). Each of these silicates was modelled with ``small'' 
and ``large'' grains.
}

\item{
Silica (SiO$_2$) coefficients were taken from Spitzer \& Kleinman 
(\cite{Spitzer}).
}

\end{itemize}

{\bf Shape and size of the dust grains:}
Astronomical dust grains likely have various shapes and structures 
(compact/fractal), which are
reflected in their different spectral signatures, complicating the spectral
analysis.  Min et al. (2003, 2005) showed that mass absorption coefficients,
calculated with a distribution of hollow spheres (DHS), give a reliable
identification of solid state components. In this work we use mass absorption
coefficients $\kappa_{i}$ from van Boekel et al.\ (\cite{Boekel05}). More 
specifically,
for the crystalline grains and silica, $\kappa_{i}$ was calculated with DHS,
while for the amorphous olivine and pyroxene grains Mie theory was applied, as
here shape effects are less significant. For a more extensive discussion of
shape and structure effects on spectral signatures, we refer to Min et al.
(2003, 2005).

In circumstellar disks, dust grains grow and get destroyed by various
processes. As a result, the dust grains do not have a single size, but a wide
range in sizes. Since the grains are not always compact, but can be fluffy as
well, the term 'volume equivalent radius', $r_{V}$ is introduced: it is the
radius of a sphere with the same material volume as the particle. This means
that fluffy particles have a much smaller $r_{V}$ than their real size. Van
Boekel et al.\ (\cite{Boekel05}) showed that the size distribution of 
grains radiating in
the 10~$\mu$m region can be represented by two grain sizes: 'small' ($r_{V}$ =
0.1~$\mu$m) and 'large' ($r_{V}$ = 1.5~$\mu$m).  It is this approach that we
follow when modelling our spectra.

{\bf Fitting procedure:}
We used absorption coefficients from van Boekel et al.\ (\cite {Boekel05})
and Bouwman (\cite{Bouwman08}), which are based on the laboratory
data cited above. For the compositional fit we perform an error analysis 
as described in Sect.\,5.1.3 of van Boekel et al. (\cite{Boekel05}), by 
minimising the expression for the reduced $\chi^2$
\begin{equation}
\chi^2_{\mathrm{red}} = \frac{1}{N_\lambda-M} \sum_{i=1}^{N_\lambda} 
\left|\frac{\mathcal{F}_\nu^\mathrm{model}(\lambda_i)-
\mathcal{F}_\nu^\mathrm{observed}(\lambda_i)}{\sigma_i}\right|^2
\end{equation}
As $\mathcal{F}_\nu^\mathrm{model}$ contains 12 free parameters,
i.e.\ the mass fractions of the five silicate emissivities for two grain 
sizes each and the continuum and dust temperature, $M$ equals 12. The 
continuum temperature was determined by fitting a 
blackbody to the continuum. $N_\lambda$ is the number of wavelength 
points $\lambda_i$, while $\sigma_i$ describes the absolute error of the 
observed flux at wavelength $\lambda_i$ and corresponds to the background 
noise derived from the data. The errors of the silicate mass 
fractions are derived from a Monte-Carlo simulation. For each spectrum
we create 100 synthetic spectra by randomly adding Gaussian noise with a 
width up to the level of the background fluctuations at each wavelength 
point. The same compositional fitting procedure is applied to each
of these spectra, resulting in slightly different mass fractions, from 
which the mean values and errors are obtained.

We caution against overinterpreting our fitting results. Indeed, significant 
systematic errors could be introduced by either an imperfect calibration or
by the limited  wavelength range and SNR, both inherent to ground-based
observations (see Juh\'asz et al.\ \cite{Juhasz} for a thorough discussion 
on this topic). Since we cannot quantify these systematic errors, we could
also not take them into account in the fitting procedure. Nevertheless, the 
results are robust in terms of distinguishing between small and large grains, 
but probably only indicative in terms of the crystalline species.

The modelling results for the emission spectra, together with the 
contribution of each dust component are shown in
\mbox{Figs.\,\ref{fig:decompo_herbig}--\ref{fig:decompo_ttauri}}. Note that, 
while we model the spectra with the 10 (grain and size) dust types and PAH as
explained above, we only plot 6 components for a better visibility. In 
particular, we added amorphous olivine and pyroxene of 0.1~$\mu$m and plot 
them as ``small amorphous silicates''. Similarly, ``large amorphous silicates''
contain the 2.0~$\mu$m pyroxene and 2.0~$\mu$m olivine grains. For silica,
forsterite and enstatite, we plot the sum of the 0.1 and 2.0~$\mu$m grains. 

When deriving in what amount the different dust species are present, the 
linear coefficients of the fit -- which are proportional to the radiating 
surface of the grains, assuming optically thin  emission -- need to be 
converted to mass. However, it is 
not possible to determine the absolute amount of mass for each species 
present, as we have no spatially resolved data to derive the particles' 
size, density or temperature distribution as a function of radius within 
the disk. Therefore, we derived mass fractions of the different species 
under the assumption that (1) the particles are spherical and (2) the 
particles have the same density. The mass fractions are meaningful to 
compare the objects in our sample and to establish the relative amount of 
processed dust we observe at 10~$\mu$m. In Table~\ref{tab:masses} we 
list the derived mass fractions, which is the ratio of the mass of a 
particular dust species and the total dust mass. Please note that for a 
given mass, small particles have a larger total emitting surface than 
large ones. Inversely, a similar amount of observed radiation will result in 
a much smaller mass when caused by small grains than if it was caused by 
larger grains. It is also important to point out that our results are only 
valid for the warm dust, which is located in the inner part of the disk 
atmosphere (r\,$<$\,10~AU) and radiates in the 10~$\mu$m region. If the disk 
atmosphere and midplane are well-mixed, the relative abundances would reveal 
the bulk composition of the inner disk, but if there is vertical settling 
the interpretation is more complex.

\begin{figure}[t]
  \centering
  \includegraphics[scale=0.58, angle=0]{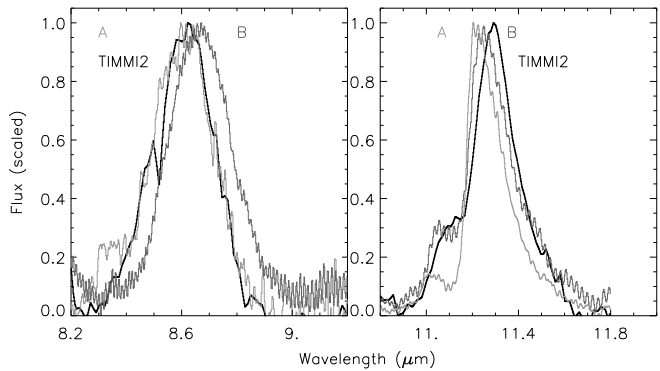}
  \caption{A comparison of the 8.6 and 11.2~$\mu$m PAH band profiles of 
           PDS 144 N (black) with the PAH classes (grey shades) as
           determined by Peeters et al.\ (\cite{Peeters02}) and van 
           Diedenhoven et al.\ (\cite{Diedenhoven}). Clearly, PDS 144 N 
           has a ``class\,A'' 8.6~$\mu$m PAH band, while the 11.2~$\mu$m 
           PAH profile is different from both class A and B. The wavelength 
           error is $\delta \lambda / \lambda \sim 10^{-3}$ for the science 
           spectra, and $\delta \lambda / \lambda \sim 10^{-4}$ for the PAH 
           class model profiles.}
  \label{fig:pds144n_2}
  \medskip
\end{figure}

\subsubsection{PAH}
\label{sect:pahs}

\begin{figure*}[t]
  \centering
  \includegraphics[scale=0.6, angle=0]{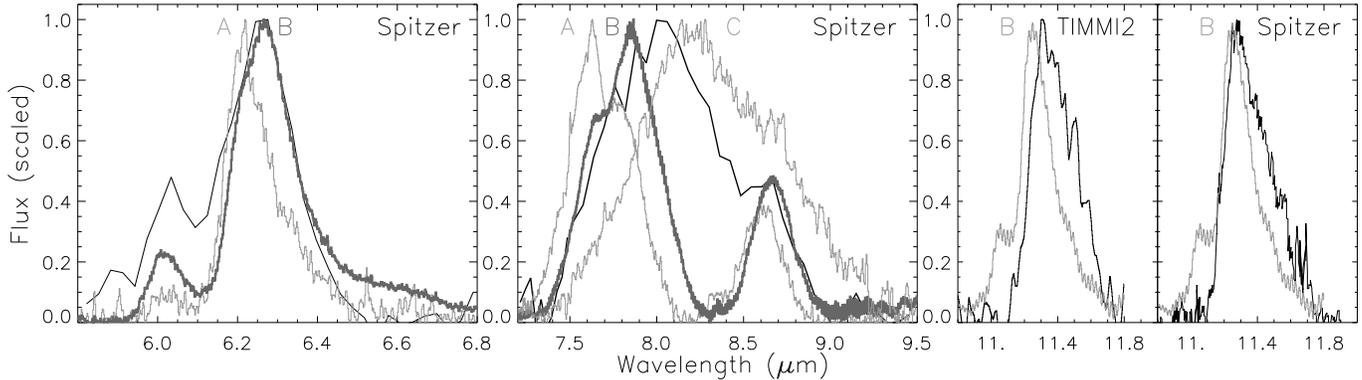}
  \caption{A comparison of the 6.2, ``7.7'', 8.6 and 11.2~$\mu$m\, PAH 
           band profiles of HD~34700 with the PAH classes as determined 
           by Peeters et al.\ (\cite{Peeters02}) and van Diedenhoven et 
           al.\ (\cite{Diedenhoven}). Observations of HD~34700 are shown 
           in thin black (both Spitzer-IRS and TIMMI2), the PAH classes are 
           shown in grey shades. Clearly, HD~34700 shows unique PAH band 
           profiles. Errors are given in the caption of 
           Fig.\,\ref{fig:pds144n_2}.}
  \label{fig:pah-hd34700}
  \bigskip
  \bigskip
\end{figure*}

Polycyclic Aromatic Hydrocarbons (PAHs) show N-band emission features at 
7.7, 8.6, 11.0, 11.2 and 12.7~$\mu$m. 
A key result in observational studies of the PAH bands is that their 
profiles show clear variations in terms of peak positions, profile shape 
and relative intensities (Peeters et al.\ \cite{Peeters04a} and references 
therein). HII regions, reflection nebulae 
and the ISM all show the same PAH band profiles (called "class A"), while 
planetary nebulae and post-AGB stars have clearly different, red-shifted 
profiles, most noticeable for the 6.2 and 7.7~$\mu$m emission bands 
(class B/C). Figs.\,\ref{fig:pds144n_2} and \ref{fig:pah-hd34700} show 
these PAH profiles for two of our targets. Note that in contrast to 
class~A, in which the PAH profiles 
in all objects are almost identical, class B comprises a large variety of 
profiles for the 6.2 and 7.7~$\mu$m PAH bands (Peeters et 
al.\ \cite{Peeters02}). The latter class is based on the spectra of 
evolved stars, where PAH formation and processing is actively taking place. 
Also HAeBe stars show large variations in their PAH profiles (Peeters et 
al.\ \cite{Peeters02}; van Diedenhoven et al.\ \cite{Diedenhoven}; Sloan 
et al.\ \cite{Sloan}). HAeBes with profiles similar to those 
found in the ISM (class A) turn out to be embedded in their parental 
cloud and PAHs from this cloud are dominating the emission. Thus, the 
initial characteristics of the PAH family entering protoplanetary disks 
of HAeBe stars are the same as in the ISM. On the other hand, those 
HAeBes exhibiting significantly different PAH profiles (class B) are 
isolated HAeBe stars and spatial studies show that the PAH emission 
originates in the disk (e.g.\ Habart et al.\ \cite{Habart04}, 
\cite{Habart06}; Geers et al.\ \cite{Geers07b}; Lagage et al.\ \cite{Lagage06};
Doucet et al.\ \cite{Doucet}), implying that PAH processing is actively
occurring in these disks. In addition, the presence of PAH emission is 
related with the disk geometry (Meeus et al.\ \cite{Meeus01}, Acke \& van den
Ancker \cite{Acke04}): 
PAH emission is observed in flaring disks where the PAH molecules 
can be excited by UV photons from the central star, while it is much weaker 
(if present) in stars with a flat disk. Also, a recent study of 
Spitzer-IRS observations of a sample of HAeBe stars and Post-AGB stars 
reveals a correlation of the position of the 7.7 and 11.2~$\mu$m PAH bands 
with the effective temperature of the central star (Sloan et 
al.\ \cite{Sloan}, \cite{Sloan07}).

\begin{figure}[t]
  \centering
  \includegraphics[scale=0.56, angle=0]{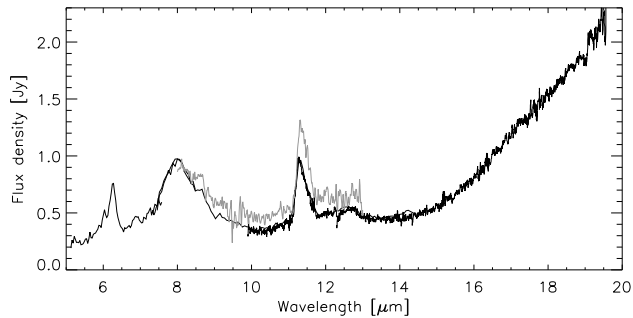}
  \caption{TIMMI2 (grey) and Spitzer-IRS (thin black and thick black) 
           observations of HD~34700 are shown. PAH emission bands are 
           clearly present on top of the dust continuum. Atmospheric
           variations may have influenced the spectrophotometric 
           accuracy of the TIMMI2 data.}
  \label{fig:t2-vs-spitzer}
  \bigskip
  \medskip
\end{figure}

\paragraph{{\bf PDS 144 N:}}

Our TIMMI2 spectrum of \object{PDS 144 N} clearly shows PAH emission at 
8.6, 11.0 and 11.2~$\mu$m (see Fig.\,\ref{fig:pds144n_1}). Comparison of 
the 8.6 and 11.2~$\mu$m PAH profiles with the different PAH classes (see 
Fig.\,\ref{fig:pds144n_2}) reveals that for \object{PDS 144 N} the 8.6~$\mu$m 
PAH band belongs to class~A and hence is similar to the ISM. In contrast, 
its 11.2~$\mu$m PAH profile is distinct from both class~A and B 
in that its peak wavelength is redshifted compared to both classes. Van 
Diedenhoven et al.\ (\cite{Diedenhoven}) examine the 11.2 \mum\, 
PAH band profile for a handful of HAeBe stars (other spectra were not 
adequate due to the presence of
crystalline silicates at this wavelength or insufficient SNR). All but one 
of these stars belong to class~A for all of their PAH bands. More recently, 
Sloan et al.\ (\cite{Sloan}, \cite{Sloan07}) reported class~B characteristics 
for four HAeBe and a T~Tauri star. In addition, these authors found a 
range of peak positions for the 11.2 \mum\, band and relate this to the 
effective temperature of the star, i.e.\ the central wavelength shifts to 
longer wavelengths as the effective temperature decreases. Tielens 
(\cite{Tielens}) pointed out that 
this relationship can also be interpreted as a step function instead of a 
continuous dependence. In this regard, the deviation from class~A seems to 
be connected to the presence of circumstellar material (Van Kerckhoven 
\cite{Kerckhoven02}; Boersma et al.\ \cite{Boersma}). However, the 
underlying cause of these variations in the 11.2~\,\mum\, band is not well 
understood (see e.g.\ van Diedenhoven et al.\ \cite{Diedenhoven}). With a 
'central wavelength' (the wavelength with half the emitted flux to either 
side, as defined by Sloan et al.\ \cite{Sloan07}) of 11.289~\mum\, and an 
effective temperature of 8750\,K, this star supports this 
relationship (note that due the asymmetry of the profile, the 'central 
wavelength' does not correspond to the peak position). Given the class A 
8.6~\mum\, PAH emission, it seems that changes happened to the present PAH 
family in this source compared to that of the ISM and that these changes 
only seem to influence the 11.2~\mum\, and not the 8.6~\mum\, PAH bands. 
Observations of the CC modes (6.2 and "7.7"~\mum\, PAH bands, which are not 
observable from the ground) and a study of PAH emission in a larger sample 
of YSOs is needed to clarify this issue.

No N-band silicate emission is seen for the northern component, which 
can be explained if the silicate grains located in the inner parts of 
the disk are either (1)~too cold, resp.\ the disk lacks a hot optically thin 
surface, or (2)~too large to emit at 10~$\mu$m, 
as shown by Meeus et al.\ (\cite{Meeus02}). Our spectrum with the prominent 
PAH features can be explained by a flared disk model (e.g.\ Acke \& van 
den Ancker \cite{Acke04}), which is also clearly recognised in the 
resolved images by Perrin et al.\ (\cite{Perrin}), who show that the
disk is optically thick in the NIR.

\paragraph{{\bf HD 34700:}}

A comparison between our TIMMI2 observation and a Spitzer-IRS observation 
of \object{HD 34700} is shown in Fig.\,\ref{fig:t2-vs-spitzer}, and 
clearly illustrates the presence of the 6.2, ``7.7'' and 11.2~$\mu$m 
PAH emission bands. A hint of the 8.6 and 12.7~$\mu$m PAH bands is
also present in the Spitzer-IRS observations while the weaker 12.0,
13.2 and 14.2~$\mu$m bands (e.g.\ Hony et al.\ \cite{Hony}) and a 
possible 15--20~$\mu$m PAH plateau/complex (e.g.\ Van Kerckhoven et 
al.\ \cite{Kerckhoven}; Peeters et al.\ \cite{Peeters04}) are not observed 
in either spectrum. 

A detailed comparison of the PAH band profiles is shown in 
Fig.\,\ref{fig:pah-hd34700}. The PAH bands exhibited by the TTS 
\object{HD 34700} are clearly distinct. In particular, its 6.2~$\mu$m band 
peaks at a similar wavelength as that of the representative of 
class B, the Red Rectangle, but also shows excess emission shortwards of 
6.1~$\mu$m and the long-wavelength tail is much weaker. A similar profile 
is observed towards the HAeBe star \object{HD 141569} (Sloan et 
al.\ \cite{Sloan}), both sources showing a very strong 6.0~$\mu$m PAH 
band (6.0/6.2 band ratio of $\sim$0.33 and 0.42 for \object{HD 141569} 
and \object{HD 34700}, respectively). Its ``7.7''\,$\mu$m complex peaks 
around 8.05~$\mu$m, redshifted compared to the Red Rectangle but, as noted 
above, class B shows a large variety and other evolved stars such as, 
e.g.\ \object{IRAS 17347-3139}, peak close to 8~$\mu$m. However, its 
profile is very broad, starting at similar wavelengths compared to that 
of the Red Rectangle, but extending up to 9~$\mu$m, thus it is more 
similar but blue-shifted to the class C profile. The latter class shows 
very weak, if any, 8.6 \mum\, emission which is also very weak in 
\object{HD 34700}. Similar broad but unique PAH profiles are also seen 
around the HAeBe stars \object{HD 141569} and \object{HD 135344} (Sloan et 
al.\ \cite{Sloan}), suggesting a smooth transition in profiles between 
class~B and C. The 11.2~$\mu$m PAH band in \object{HD 34700} peaks at a 
longer wavelength and also shows excess emission longwards, when compared to 
the profile of the Red Rectangle. Note that the 11.0~$\mu$m PAH band, usually 
attributed to ionised PAHs (Hudgins \& Allamandola \cite{Hudgins}; Hony et 
al.\ \cite{Hony}), is not observed in \object{\mbox{HD 34700}}. The TIMMI2 
and Spitzer-IRS profiles are rather similar with only small differences at 
the blue wing of the 11.2~$\mu$m band. The 'central wavelengths' (the 
wavelength with half the emitted flux on either side, as defined by Sloan 
et al.\ \cite{Sloan07}) are 8.06, 
11.35 and 11.37~\mum\, for respectively the 7.7~\mum\, complex, the 
11.2~\mum\, band in the IRS-SH observations and the 11.2~\mum\, band in the 
TIMMI2 observations. While this is consistent with the results of Sloan et
al.\ (\cite{Sloan07}) for the 11.2~\mum\, band, the 'central wavelength' of 
the 7.7~\mum\, complex is considerably shorter for an effective temperature of
6000\,K.

Summarising, the PAH band profiles of the T~Tauri star 
\object{\mbox{HD 34700}} are clearly unique and show that PAH processing 
has occurred in this source. This is similar to PAH observations towards 
isolated HAeBe stars and a few other T Tauri stars and suggests that similar 
PAH processing occurs in protoplanetary disks around low-mass and 
intermediate-mass stars.

Just like for \object{PDS 144 N}, the absence of silicate emission in 
\mbox{\object{HD 34700}} can be explained if the silicate grains located 
in the inner parts of the disk are either too cold or too large to emit at 
10~$\mu$m. The shallow sub-mm slope of the SED in Fig.\,\ref{fig:sed_ttauri} 
suggests grain growth for 
those particles located in the outer disk and emitting at (sub-)mm 
wavelengths, but this does not necessarily imply the same evolution in the
more inward disk regions. Although much less IR 
excess is seen in this SED, the prominent PAH emission suggests that the 
disk is still flared to some extent, and not self-shadowed. Alternatively,
and taking into account that this system is a spectroscopic binary, the 
lack of warm dust could also be explained by an inner hole in a circumbinary 
disk, while the PAHs may be located further outwards and still emit at 
10~$\mu$m.

\section{Results}
\label{sect:discussion}

\subsection{Relations between stellar, disk and silicate properties}

In the following we discuss the silicate shapes and study the silicate 
properties as a function of (1)~stellar 
luminosity -- between $\sim$0.5 and $\sim$500 L$_{\odot}$, (2)~binarity -- 
with 
separations from approximately 0.1 to 5000 AU, and (3)~infrared SED slope. 
To obtain more meaningful relations, we also consider the targets presented in 
Paper\,I and~II. Apart from the three debris
disks \object{HD 113766}, \object{HD 123356} and \mbox{\object{HD 172555}}, 
all other targets are PMS objects. We will distinguish both disk types in the 
following discussion, when required.

\begin{table}[t]
  \begin{center}
  \caption{Labels and luminosities (with errors when they were given in the
           literature) used in 
           Figs.\,\ref{fig:shape-strength}--\ref{fig:strength-slope}.}
  \setlength\tabcolsep{11.3pt}
  \begin{tabular}{rlcr}
    \hline
    \hline
    Label &  Target      &   $L/L_{\sun}$               &  Ref. \\
    \hline
    \noalign{\medskip}
      1   &  HD~72106 B  &   9.9$_{- 5.6}^{+  12.7}$    &  (1)  
      \medskip \\
      2   &  HD~98800 B  &   0.58$_{- 0.11}^{+  0.11}$  &  (2)  
      \medskip \\
      3   &  HD~113766 A &   4.4$_{- 0.4}^{+  0.4}$     &  (3)  
      \medskip \\
      4   &  HD~123356 B &   ...                        &  (4)  
      \medskip \\
      5   &  HD~143006   &   0.8                        &  (5)  
      \medskip \\
      6   &  HD~172555   &   $>$16                      &  (6)  
      \medskip \\
      7   &  HD~190073   &   83                         &  (7)  
      \medskip \\
      8   &  HD~319139   &   0.49$_{- 0.06}^{+  0.06}$  +
                             0.33$_{- 0.04}^{+  0.04}$  &  (8)  
      \medskip \\
      9   &  CD-43~344   &   ...                        &  (9)  
      \medskip \\
     10   &  FU~Ori A    &   466                        &  (10) 
      \medskip \\
     11   &  KK~Oph A    &   20$_{- 4}^{+  5}$          &  (11) 
      \medskip \\
     12   &  MP~Mus      &   $>$0.45                    &  (6)  
      \medskip \\
     13   &  PDS~144~S   &   $>$16                      &  (6)
      \smallskip \\
    \hline
  \end{tabular}
  \label{table:labels}
  \end{center}
  \vspace{0.15cm}
           References: (1)~Folsom et al.\ (\cite{Folsom}), 
           (2)~Prato et al.\ (\cite{Prato}), (3)~Lisse et 
           al.\ (\cite{Lisse}), (4)~neither distance nor the spectral type 
           are certain, (5)~Natta et al.\ (\cite{Natta}), (6)~lower limit 
           when assuming a ZAMS luminosity for the given spectral types, 
           (7)~Acke et al.\ (\cite{Acke05}), (8)~Quast et 
           al.\ (\cite{Quast}), (9)~no luminosity class is known, 
           (10)~Green et al.\ (\cite{Green}), (11)~adapted by Carmona et 
           al.\ (\cite{Carmona}).
  \bigskip
  \medskip
\end{table}

\paragraph{{\bf Silicate shape in function of peak strength:}}

The silicate mass fractions (Table~\ref{tab:masses}) show no clear 
dependence on stellar parameters. As a function of age, we see a decrease 
of amorphous silicates and an increase of crystalline forsterite and 
enstatite, in agreement with common expectations, but no trend of grain
growth with disk lifetime. In order to have the discussion on more solid 
grounds, we use spectral indices as they are more robust, since they are 
model independent. 
The silicate shape, i.e.\ the ratio of the fluxes at 11.3 and 9.8~$\mu$m in 
the continuum normalised spectra, constitutes a good probe to study 
trends and correlations, and can be seen as a direct indicator for the grain 
size (Bouwman et al.\ \cite{Bouwman01}, van Boekel et al.\ \cite{Boekel03}).

\begin{figure}[t]
  \centering
  \includegraphics[scale=0.49, angle=0]{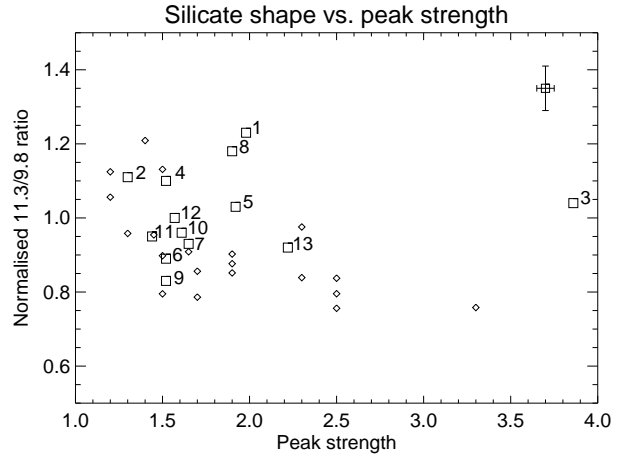}
  \caption{Silicate shape (the 11.3/9.8 ratio) in function of peak strength. 
           While HD~113766 (data point '3') stands out due to its secondary 
           generation dust, and sources of very pristine dust are missing 
           in our sample, the data follow the same trend as found in 
           previous studies. The target number labels are explained in 
           Table~\ref{table:labels}. Typical error bars are given in the top 
           right corner. For comparison, data from
           Sicilia-Aguilar et al.\ (\cite{Sicilia}) is included (small 
           diamonds).}
  \label{fig:shape-strength}
  \bigskip
  \medskip
\end{figure}

The silicate shape in function of the peak strength is shown in 
Fig.\,\ref{fig:shape-strength}. To make our results comparable to previous 
studies, we follow the same definition of the silicate peak strength and 
shape as given in Sect.\,4.2 of van Boekel et al.\ (\cite{Boekel05}). In 
a weighted linear fit with relation $y = a + bx$, where $x$ is the 
feature strength and $y$ the flux ratio, values for $b$ between
$-0.28 \leq b \leq -0.16$ have been reported (van Boekel et 
al.\ \cite{Boekel03}; Przygodda et al.\ \cite{Przygodda}; Kessler-Silacci 
et al.\ \cite{Kessler07}). We cannot give a mathematical relation for our 
sample, as it is lacking sources of very pristine dust (which would 
populate the middle bottom in Fig.\,\ref{fig:shape-strength}). In order 
to compare with a larger sample, we added data from Sicilia-Aguilar 
et al.\ (\cite{Sicilia}). The Vega-type \object{HD 113766} (label number 3) 
has an isolated position in our plot, due to its second generation dust and 
its double peak in the spectrum. Besides this object, our sample of PMS disks 
agrees qualitatively well with the correlations found in the cited works: 
a decrease of the peak strength correlates with increasing dust 
processing (i.e.\ with a larger 11.3/9.8 ratio).

\paragraph{{\bf Grain processing as a function of stellar luminosity:}}

\begin{figure}[t]
  \centering
  \includegraphics[scale=0.49, angle=0]{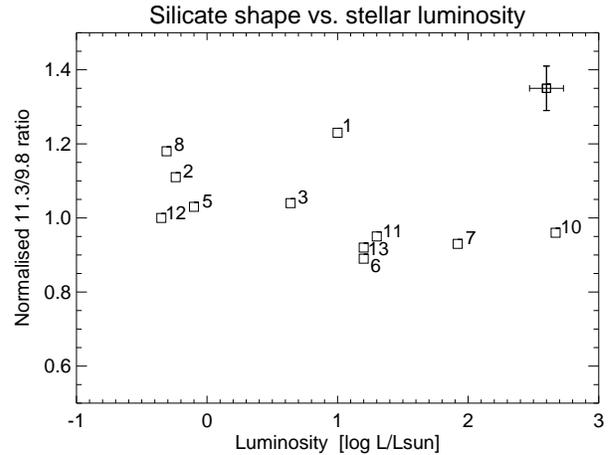}
  \caption{Silicate shape in function of stellar luminosity. Objects with 
           uncertain luminosity are excluded in this plot. Typical error 
           bars are shown in the top right corner. The luminosities are 
           listed in Table~\ref{table:labels}.}
  \label{fig:shape-lum}
  \bigskip
  \medskip
\end{figure}

No correlation of the crystalline mass fractions (Table~\ref{tab:masses}) with 
stellar luminosity is found in our sample, nor for the derived grain size. 
Kessler-Silacci et al.\ (\cite{Kessler07}) have 
shown a correlation of the silicate shape with stellar luminosity, in form 
of an observed larger grain size (as witnessed by the 
11.3/9.8 ratio) with lower luminosity. This trend can be understood when
considering that the 10~$\mu$m silicate emission region lies further inward 
for stars with lower luminosity, assuming that the grain size distribution 
with radius is the same in all disks, and grain growth is most efficient in 
the inner, denser regions. Our sample is much smaller than that presented 
by Kessler-Silacci et al.\ (\cite{Kessler07}), thus we cannot give a 
mathematical relation due to the scatter of our data points, but 
Fig.\,\ref{fig:shape-lum} suggests a similar trend with luminosity.

\paragraph{{\bf Influence of companions:}}

We see no correlation of the degree of dust processing with stellar 
binarity in our sample, i.e.\ whether the time spans of dust processing 
and disk life time in binary systems would differ from those of isolated 
stars of the same age. 75\% of our targets are binaries or multiple systems. 
These include systems without N-band silicate emission (e.g.\ HD~3003, 
HD~80951, HD~155555, HD~181296), objects with mainly unprocessed dust 
(FU~Ori) or large dust grains (KK~Oph, HD~319139), and sources with all 
types of silicates present (especially HD~123356). 

For most binaries in our sample the separation (cf.\ Table~\ref{table:targets})
is around 200~AU -- apart from \object{PDS 144} with $\sim$5000~AU, while 
dust in very different processing stages is found. In case of the three
spectroscopic binary (SB) systems, silicate dust is either seen in small
amounts or is absent, which can be explained by the removal of dust inside the 
circumbinary disk. Recently, Pascucci et al.\ (\cite{Pascucci}) have shown 
that companions at
separations of tens of AU have no impact on the dust evolution. It would be 
interesting to study this also for a well defined target sample with closer 
companions.

\paragraph{{\bf Silicate strength in function of SED slope:}}

\begin{figure}[t]
  \centering
  \includegraphics[scale=0.49, angle=0]{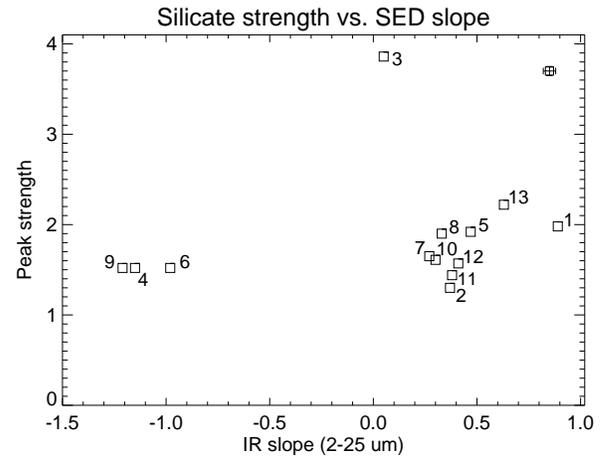}
  \caption{Silicate peak strength versus infrared
           \mbox{2--25}~$\mu$m SED slope (\mbox{$d$ log $F_{\lambda}$} /
           $d$ log $\lambda$). HD~113766 (data point '3') has a particular 
           position due to its large amounts of secondary generation dust.
           See Table~\ref{table:labels} for the number labels. Typical 
           error bars are plotted in the top right corner.}
  \label{fig:strength-slope}
  \bigskip
  \smallskip
\end{figure}

With ongoing grain growth and subsequent settling, an initially flared 
PMS disk is thought to evolve into a flat disk (Dullemond \& Dominik 
\cite{Dullemond}), accompanied by a change in the slope of the SED: the
initially positive \mbox{2--25~$\mu$m} slope will decrease. During 
the evolution into a debris disk, the IR emission from the inner disk 
decreases when the disk is cleared from inside out. This will result
in a steeper negative SED slope with progressing evolution into a debris
disk. When the dust has nearly completely disappeared, the slope will 
approach that of a blackbody with the stellar temperature. 
A dependence of silicate emission on inner disk evolution would be 
recognisable as a correlation between silicate strength and SED slope. 
There is a tendency that more flared (and therefore redder) 
disks have stronger silicate features. We used an average extinction curve 
(Savage \& Mathis \cite{Savage}) with R\,=\,3.1 and the $A_v$ in
Table~\ref{table:targets} to calculate the de-reddening of the SEDs. The 
slope between \mbox{2--25~$\mu$m} is calculated as
\begin{equation}
\alpha_{2-25} = 
\frac{\log F_{25} - \log F_{2}}{\log \lambda_{25} - \log \lambda_{2}},
\end{equation}
where $\mathrm{F}$ is the flux in Jansky at the indicated wavelengths. 
Sicilia-Aguilar et al.\ (\cite{Sicilia}) noticed a very weak to absent 
correlation between silicate strength and SED slope, with which our data 
agree (see Fig.\,\ref{fig:strength-slope}). Vega-type and pre-main 
sequence stars can be clearly distinguished by their different slopes, 
but a convincing dependence on the silicate strength is not seen. 
\object{CD-43 344} has an unusual position for a PMS source in this plot,
but as noted earlier, its classification is not very firm.

\subsection{Discussion}
\label{sect:summary}

We have presented new N-band spectra of 15 PMS and Vega-type stars. Besides 
\object{FU Ori}, these spectra were either previously unpublished or not 
yet available in a good quality to permit the analysis of the PAH bands 
or a dust decomposition of the silicate emission. HD~155555~AB is 
the only T~Tauri star in our sample without an IR excess, while four of 
the five Vega-like stars only show photospheric emission at 10~$\mu$m, 
with excess emission only at longer wavelengths. For 
the remaining targets we analysed their silicate and PAH features:

\begin{itemize}

\item{
{\bf Silicate dust}: Unprocessed dust dominates the circumstellar matter 
around \object{FU Ori}, \object{HD 123356}, \object{HD 143006} and
\mbox{\object{CD-43 344}}. \object{HD 123356} is considered a 
Vega-type object, and we show for the first time that the dust
emission arises from the secondary. No firm age estimation is available 
and we do not rule out that this system may be younger than currently 
believed. The N-band profile of \object{\mbox{HD 123356 B}} is rather
unusual, dominated by small amorphous silicates with all types of 
processed dust present as well. High-resolution NIR and MIR studies are 
required to learn more about this system. Similarly, the classification of 
\mbox{\object{CD-43 344}} as a T~Tauri star is not unambiguous. No distance, 
luminosity class or age estimation is available. Its dust properties could 
also be explained by an evolved Mira-type star. Optical 
spectroscopy is required to determine the stellar class.

Large amorphous silicates are the main dust component towards
\object{HD 190073}, \object{HD 319139}, \object{KK Oph}, and 
\object{\mbox{PDS 144 S}}. No source in our sample shows highly processed 
dust, i.e.\ a spectrum dominated by crystalline silicates, as we found 
for some targets presented in Paper\,I and~II (e.g.\ the Herbig~Ae star 
\object{HD 72106 B} or the Vega-type \mbox{\object{HD 113766 A}}). In agreement
with previous studies, we find no correlation of dust processing with stellar 
age in our samples. Although \object{HD 181296} and  
\object{\mbox{HD 172555}} have a similar age and spectral type, the first
source has no excess emission in the N-band, while the last, presented 
in Paper\,II, shows dust in an early stage of processing. This could be 
explained by a cleared inner disk region of different sizes, with the disk 
region sensitive to N-band observations ($<$10~AU) already cleared in 
\object{HD 181296}. 

We see no correlation of the dust evolution with binarity in our sample,
i.e.\ whether the dust processing in binary systems is retarded or
accelerated when compared to isolated stars.
}

\item{
{\bf PAH emission:}
PAH bands are observed towards the HAeBe star \object{PDS 144 N} and the 
TTS \object{HD 34700}. Their presence is often observed in HAeBe 
stars and commonly explained with the excitation of PAH molecules by 
UV photons. However, Li \& Draine (\cite{Li}) have shown that PAH 
molecules around cool stars can also be excited. Laboratory experiments 
by Mattioda et al.\ (\cite{Mattioda}) revealed electronic absorption bands 
of PAH ions up to the NIR. Thus, PAH emission can also be found in 
UV-poor environments. In 5--35~$\mu$m spectra of a sample of 38 TTS, Geers et
al.\ (\cite{Geers06}) found PAHs for 3 targets (with possible further 
tentative detections). In their sample, \object{T Cha} of spectral type G8 
represents the lowest mass source with PAH emission. Furlan et 
al.\ (\cite{Furlan}) found 2 cool stars with PAH emission in 5--36~$\mu$m 
spectra of 111 targets in the Taurus-Auriga star-forming region. They 
show \object{UX Tau A} as the coolest TTS with PAH up to now, having
a spectral type K5, but the PAH feature in this source is rather marginal. 
The very latest-type PAH emitting TTS presently is \object{IRS 48} (Geers 
et al.\ \cite{Geers07a}), while these authors point out that its apparent 
spectral type (M0) has been highly variable in the last decade, possibly due to
a FUOR-type accretion event.

The PAH profiles of \object{HD 34700} show that PAH processing occurs 
similarly to PAHs observed towards isolated HAeBe stars, suggesting that 
similar PAH processing exists in the protoplanetary disks around low-mass 
and intermediate-mass stars. The case is less clear for the HAeBe star 
\mbox{\object{PDS 144 N}}. Although the 8.6~$\mu$m PAH band profile is 
identical to that in the ISM, the 11.2~$\mu$m PAH band is clearly distinct.
}

\end{itemize}

\object{PDS 144} constitutes a very interesting system, with prominent 
PAH bands towards the Northern source, while the Southern component is
dominated by silicate emission. Although it is commonly believed that both 
objects form a binary, this remains to be proven. Perrin et al.\ (private 
communication) recently obtained optical spectra to determine more accurate 
spectral types and an improved distance estimate, to give a more solid 
answer to this question.

\section{Conclusions}
\label{sect:conclusions}

We presented new N-band spectroscopy and modelled the 10~$\mu$m 
emission features, in order to characterise the circumstellar dust in our
sample. We also discovered previously unknown circumstellar 
disks. Our main conclusions are:

\begin{enumerate}

\item{
We characterised the PAH emission in \object{PDS 144 N}
and \mbox{\object{HD 34700}}. \object{PDS 144 N} exhibits mixed PAH class 
characteristics, while \object{HD 34700} clearly belongs to class~B. The 
latter points towards processing of PAHs, suggesting a similar PAH 
evolution in the disks of Herbig and T~Tauri stars.
}

\item{
The emission spectrum of \object{FU Ori} resembles that of 
\mbox{\object{BBW 76}} in Paper\,I, while the FUORs \object{V346 Nor}, 
\mbox{\object{V883 Ori}} and \object{Z CMa} of Paper\,I show silicate 
absorption features. This could be an effect of the disk orientation (pole-on 
vs.\ face-on) or represent the two FUOR categories defined by Quanz et 
al.\ (\cite{Quanz07}).
}

\item{
We showed the first N-band spectrum of \object{CD-43 344}, a star that 
hosts only pristine silicates and was poorly studied. However,
its luminosity and stellar class are rather uncertain.
}

\item{
Silicate dust with mixed properties (sizes and crystallinity) is found towards 
\mbox{\object{HD 123356 B}}, while high-resolution imaging is required for 
a further understanding of this particular system.
}

\item{
Within our sensitivities no N-band silicate emission is seen for the T~Tauri 
star HD~155555~AB and the Vega-type objects \mbox{\object{$\eta$ 
Corvi}}, \mbox{\object{HD 3003}}, \object{HD 80951} and \object{HD 181296}.
}

\item{
We failed to detect \object{HD 158153} and \object{SAO 185668} despite of
their alleged high IRAS flux and sufficient integration time. We attribute
this to other sources that could be present in the IRAS 
beam, causing erroneous fluxes for our target. In fact, there are quite a lot 
of other stars to which an IRAS flux had been attributed in the 1990s, but 
which later got revised, when higher spatial resolution was possible in the IR.
}

\item{
We find no dependence of dust processing with age or stellar binarity, for
a companion separation of typically $>$100~AU.
}

\end{enumerate}

\begin{acknowledgement}

GM acknowledges financial support by the Deut\-sche Forschungsgemeinschaft 
(DFG) under grants ME2061/3-1 and /3-2. We would like to thank Alexis 
Brandeker for the analysis and discussion of the photometric VISIR data of 
$\eta$ Corvi, and Aurora Sicilia-Aguilar for providing the additional data 
plotted in Fig.\,\ref{fig:shape-strength}. 
This work made use of the SIMBAD astronomical database. We thank the La 
Silla staff and telescope operators for support during the observations.

\end{acknowledgement}

\newpage

\begin{appendix}

\section{Comments on individual sources}

\subsection{FU Ori stars}
\label{sect:fuor}

FU Orionis objects (FUORs) show a variability which is characterised by 
dramatic outbursts in optical light, followed by a fading phase which can 
last decades. The origin of this phenomenon is not clear but it is often 
associated with enhanced accretion in low-mass pre-main sequence stars 
(Hartmann \& Kenyon \cite{Hartmann}). 

\paragraph{{\bf FU Ori}} is the prototype of the FU~Orionis phenomenon. 
While the SIMBAD database lists a spectral type of G3Iavar, Kenyon et 
al.\ (\cite{Kenyon00}) changed it to G0II. In contrast to Hanner et 
al.\ (\cite{Hanner}), who find an increasing MIR flux shortward of 
10~$\mu$m, the spectra of Schegerer et al.\ (\cite{Schegerer}) and Quanz 
et al.\ (\cite{Quanz}) agree very well with ours, indicating that this 
object was not variable in the MIR between Dec.\ 2002 and Jan.\ 2006 (or
at least that a possible period with variability was not observed). No 
Kurucz model was plotted in the SED of \object{FU Ori} as explained in 
Sect.\,\ref{sect:sed}, but it is recognisable that the total IR excess 
of this source is larger than for all other targets in this work. The 
analysis of the silicate emission feature shows that it is dominated 
by small amorphous silicates. Beside some forsterite, no other 
dust species is found for this target in relevant amounts. 

When compared to the four FUOR spectra shown in Paper\,I, we see that 
\object{FU Ori} agrees well with the spectral profile of 
\object{\mbox{BBW 76}}, while the other 3 FUORs \object{V346 Nor}, 
\object{V883 Ori} and \object{Z CMa} show a very different feature, 
as it is in absorption. The appearance of FUOR MIR spectra 
could be the result of a geometrical effect: pole-on (emission feature) versus 
edge-on (absorption feature) view of the disk. However, given the large number
of FUORs with silicate absorption features, this difference cannot be 
explained by disk orientation alone. Therefore, Quanz et al.\ (\cite{Quanz07}) 
defined two categories of FUORs: (1) objects where the silicate feature is in 
absorption are likely still embedded in a circumstellar envelope, while (2)
objects where the silicate band is in emission are thought to have largely 
shed their envelope, and the feature arises in the surface layer of their 
accretion disks.

\subsection{Herbig Ae/Be objects}

\paragraph{{\bf HD 190073}}

has been discussed in the scientific literature 
since more than 70 years with various classifications. Being previously 
considered a post-main sequence giant, van den Ancker et 
al.\ (\cite{Ancker}) describe \object{HD 190073} as a Herbig Ae/Be star, 
what is now generally accepted. Considering its large distance, for which 
only a lower limit is known due to an imprecise Hipparcos parallax (van 
den Ancker et al. \cite{Ancker}), this unusual object must be much more 
luminous than other Herbig stars of similar spectral type. The optical 
spectrum is also quite peculiar (Pogodin et al.\ \cite{Pogodin}).

The SED shows an IR excess which begins in the NIR and rises to the 
far-IR. The decomposition of the N-band silicate feature shows a strong
dominance of large amorphous silicates. Silica and enstatite are present 
in much smaller amounts, while no forsterite is found.

\paragraph{{\bf KK Oph}}

is a PMS binary with a 1.6$\arcsec$ separation 
(e.g.\ Leinert et al.\ \cite{Leinert97}) that shows brightness variations 
of the UX Orionis type (reddens when becoming fainter). Very controversial 
spectral types have been determined (between Be and F0) and the 
system was repeatedly suggested to constitute a Herbig Ae star with a 
K-type classical T~Tauri companion (e.g.\ Leinert et al.\ \cite{Leinert97}; 
Hernandez et al.\ \cite{Hernandez}; Herbig \cite{Herbig05}). Recently, 
Carmona et al.\ (\cite{Carmona}) derived a spectral type A6Ve for the 
primary and G6Ve for the companion. Various 
estimates put this target at a distance between 160 and 310~pc, with the 
lower limit being most frequently cited. 

The N-band spectrum discussed here was already shown in Leinert et 
al.\ (\cite{Leinert04}), but the silicate feature was not yet analysed
in terms of dust composition. In the TIMMI2 images we see one isolated 
source, most likely just the HAe star itself, but we note that the MIR 
sensitivities were not good during that night. The ESO/ST-ECF Science 
Archive hosts another TIMMI2 spectrum of \object{KK Oph}, which is described 
in Geers et al.\ (\cite{Geers05}). These two spectra differ by about 
10\% in the 8~$\mu$m flux, while being in agreement longward of 10~$\mu$m. 
Both spectra, as well as the standard stars, were observed at an airmass 
close to 1.0. The distance in time between object and closest
standard star was approximately one hour in both cases. We selected the 
first mentioned spectrum, since it matches closer the TIMMI2 N1 photometry 
from the Science Archive.

The decomposition into dust components shows quite some diversity. Large 
amorphous silicates are dominating, but silica and forsterite are also 
found in significant amounts, while enstatite is not seen.

\paragraph{{\bf PDS 144 N:}}

Torres et al.\ (\cite{Torres95}) discovered during the Pico dos Dias 
Survey the binarity of this Herbig Ae/Be star with a 5$\arcsec$ 
separation. Spectral types of both components, given in 
Table~\ref{table:targets}, were determined by Vieira et 
al.\ (\cite{Vieira}). Recently, Perrin et al.\ (\cite{Perrin}) 
presented comprehensive multi-wavelength observations and discussed the 
possible distance to \object{PDS 144}, proposing a value of 1000 $\pm$ 
200~pc, while previous estimates gave $d$ = 140--2000~pc. Although it is 
commonly believed that both stars form a binary system, this has not yet 
been explicitly proven. The dominance of PAHs in \object{PDS 144 N} can 
be recognised from the NIR images in Perrin et al.\ (\cite{Perrin}), 
showing an optically thick, edge-on flared disk, where the emission in the
3.3~$\mu$m band clearly stands out in intensity. Until now, N-band spectra 
of both components have not yet been published.

\paragraph{{\bf PDS 144 S:}}

Resolved circumstellar emission from the more luminous component 
\object{\mbox{PDS 144 S}} was not detected in our and previous N-band 
images, but has recently been found by Perrin et al.\ (private 
communication) at 18~$\mu$m (0.57$"$ diameter). In the N-band spectrum we 
see silicate features of mainly large amorphous silicates, in addition to 
small amorphous grains and silica. Forsterite and enstatite are absent, 
pointing to unprocessed dust, while small amounts of PAHs might be present 
as well. 
We note that our spectrum was taken during a night with strong variations 
in the MIR transparency, but great care was taken during its calibration 
by applying a close standard star at similar airmass observed immediately 
after the science exposure.

Both components of \object{PDS 144} have a similar spectral type and, 
assuming they were formed together, a same age, but show  
different emission features: PAH versus silicates. The presence/absence of PAH
bands could be explained by the disk geometry, as we would observe much 
less PAH emission in a flat, self-shadowed disk, which prevents the UV 
photons from reaching the PAH molecules (e.g.\ Dullemond \& 
Dominik \cite{Dullemond}). On the other hand, we do not want to exclude
the trivial explanation that the amounts of PAH in both disks could be 
different, despite the similarity of both sources.

\subsection{T~Tauri targets}

\paragraph{{\bf HD 34700}} 

constitutes a quadruple T~Tauri system. For the inner, 
short-period (P\,$\sim$\,23.5 days) spectroscopic binary,  Torres 
(\cite{Torres04}) had performed an orbital solution and showed that both 
stars are of similar mass and luminosity. Sterzik et al.\ (\cite{Sterzik}) 
found two more, faint, late-type components of PMS nature. Despite 
several indicators for stellar youth (IR excess, X-ray emission, 
H$\alpha$ emission) the precise age has not yet been established, since 
the Hipparcos parallax is very uncertain (0.86 $\pm$ 1.84~mas). 

Sub-mm observations of \object{HD 34700} were reported by, e.g., Syl\-vester 
et al.\ (\cite{Sylvester01}) and Sheret et al.\ (\cite{Sheret}). An N-band 
spectrum was shown in Sylvester \& Mannings (\cite{Sylvester00}), 
but its low SNR does not permit a detailed study. Our TIMMI2 spectrum 
is also of higher quality than the ISO-SWS and ISO-PHOT data of Acke \& 
van den Ancker (\cite{Acke04}). As can be seen from the Kurucz model, 
excess emission starts around 5~$\mu$m and becomes dominant at MIR to mm 
wavelengths. A PAH band feature at 3.3~$\mu$m was shown by Smith et 
al.\ (\cite{Smith}).

\paragraph{{\bf HD 143006:}}

As for several other PMS objects also this star was previously erroneously 
classified as pre-planetary nebula, but now it is commonly considered as 
a TTS (see, e.g., Mamajek et al.\ \cite{Mamajek}). Contrary to the 
assumption of a disk-like distribution of the 
circumstellar dust, Dent et al.\ (\cite{Dent}) found single-peaked, narrow 
CO lines, which are more consistent with emission from an envelope  
than coming from a disk.

Our SED shows a large NIR excess attributable to hot dust emission. Such a 
large NIR excess is often associated with polarised emission from the 
circumstellar 
matter (Yudin \cite{Yudin}), but Hales et al.\ (\cite{Hales}) did not find 
any polarised emission and conclude that the disk may be too small to be
resolved in their UKIRT observations.

The decomposition of the N-band spectrum reveals dust at a very early 
stage of processing. Small amorphous silicates dominate the N-band feature, 
next to smaller amounts of silica, enstatite and forsterite. Moderate 
PAH emission is seen on top of the silicate feature, causing small peaks 
in the spectrum between 8-9~$\mu$m and at 11.2~$\mu$m.

\paragraph{{\bf HD 155555}}

(alias V824 Arae) is a member of the $\beta$~Pic 
moving group (Torres et al.\ \cite{Torres06}) and consists of a G5IV star 
in a short-period orbit (P\,$\sim$\,1.68 days) with a K0IV-V spectroscopic
companion (e.g.\ Strassmeier \& Rice 
\cite{Strassmeier}). Another M4.5 visual companion is located 33$\arcsec$ 
away. The system was considered to be a RS CVn type, however, Pasquini et 
al.\ (1991) proposed it to be a TTS. Both 
classifications are still being used. A single source is seen in our 
TIMMI2 images, while we were pointing at the position of the A and B 
component. The C component was just outside the field of view, which 
explains why IRAS obtained a higher 12~$\mu$m flux than observed in our 
data. Within our sample of T~Tauri stars, HD~155555~AB constitutes 
the only target without detected excess emission above the stellar 
photosphere.

\paragraph{{\bf HD 319139:}}

The spectroscopic binary TTS and member of the $\beta$~Pic moving group 
(Torres et 
al.\ \cite{Torres06}) \object{\mbox{HD 319139}} (alias V4046 Sgr) is known 
to be an 
emission-line star since the year 1938. For a summary of its "early 
history" see, e.g., Byrne (\cite{Byrne}). Hutchinson et 
al.\ (\cite{Hutchinson}) performed photometry from the UV to the 
mid-IR Q-band. From the absence of a NIR excess, Jensen \& Mathieu 
(\cite{Jensen}) concluded there is an inner hole in its circumbinary disk. 
Quast et al.\ (\cite{Quast}) refined the orbital period to 2.421 days and 
determined the spectral types of both components to K5Ve and K7Ve. All 
before mentioned authors give distance estimates ranging 
between 42--83~pc. We refer in this work to 83~pc (Quast et 
al.\ \cite{Quast}), as this is the most recent estimate.

Our N11.9 flux (central wavelength 11.6~$\mu$m) of \object{HD 319139} lies 
about 20\% below the corresponding IRAS flux, but it matches very well the 
11.5~$\mu$m flux determined by Hutchinson et al.\ (\cite{Hutchinson}). 
We see a rising dust continuum, on top of which large amorphous silicates
dominate the emission feature, followed by small amorphous grains
and some forsterite and silica. The recently published N-band
spectrum by Honda et al.\ (\cite{Honda}), part of a sample of 30 low-mass 
stars observed with COMICS at the Subaru Telescope, agrees with our data.

\paragraph{{\bf CD-43 344:}}

Mannings \& Barlow (\cite{Mannings98}) proposed the F6V star 
\object{HD 7151} as a debris disk candidate, by cross-correlating the 
``Michigan Catalog of Two-dimensional Spectral Types for the HD Stars'' 
with the ``IRAS Faint Source Survey Catalog'', associating 
\object{IRAS F01089-4257} with this star. Sylvester \& Mannings 
(\cite{Sylvester00}), however, showed that the more closely located M2 
star \object{\mbox{CPD-43 142}}, primarily known as \object{CD-43 344}, 
can better explain the emission of this IRAS FSC source. Despite a 
significant MIR excess, this target was not yet subject of any further 
studies. Here we consider this target to be a TTS, based on its 
spectral type and pristine status of dust processing, but this 
classification is not a unique solution as, e.g., an evolved Mira-type 
star could show the same dust emission. Currently the distance to 
\object{CD-43 344} and its luminosity class are undetermined. 

The 2MASS fluxes in the SED might be influenced by other sources, 
since the attributed NIR excess seems extraordinarily
high. In the MIR, however, the target appears isolated. The decomposition 
of the silicate emission feature is dominated by small amorphous silicates 
with a less prominent contribution of larger grains. We did not find 
crystalline dust.

\subsection{Vega-type stars}

\paragraph{{\bf HD 3003}}

is a proposed member of the Tucana/Horologium 
Association (Zuckerman \& Song \cite{Zuckerman04}). In the ``Catalog 
of Components of Double \& Multiple stars'' (Dommanget \& Nys 
\cite{Dommanget}) it is listed as a binary system with 0.1$\arcsec$ 
separation and almost equal visual brightness of the components. Since this 
measurement is listed as being obtained in 1925, we are cautious 
about this information. In our TIMMI2 data (with 
$\sim$0.2$\arcsec$ pixel resolution) only one isolated point-like 
emission source is seen. Song et al.\ (\cite{Song01}) give for this star 
an age estimation of 50~Myr with an upper limit of 247~Myr.
The SED for this target shows no IR excess emission up to the N-band,
with little excess seen at 25~$\mu$m.

\paragraph{{\bf HD 80951}}

constitutes a quadruple system (Dommanget \& Nys 
\cite{Dommanget}). Oudmaijer et al.\ (\cite{Oudmaijer}), as well as
Mannings \& Barlow (\cite{Mannings98}) characterised it as a system
with excess emission in the IRAS passbands, but no further indication 
of circumstellar matter has been described in the literature.
Our TIMMI2 N-band spectrum is consistent with photospheric emission.
From the SED we see an IR excess starting around 25~$\mu$m.

\paragraph{{\bf HD 109085:}}

Stencel \& Backman (\cite{Stencel}) and later Mannings \& Barlow 
(\cite{Mannings98}) characterised \object{HD 109085} (alias $\eta$~Corvi) 
as a Vega-type object. At 850~$\mu$m, a disk is found and resolved with a 
size of $\sim$100~AU (Wyatt et al.\ \cite{Wyatt}). As shown by the SED of this 
target, excess emission is insignificant at 10~$\mu$m and starts longward 
of $\sim$17~$\mu$m, extending into the sub-mm regime. Our TIMMI2 N-band 
spectrum appears stellar. In Spitzer IRS data, Chen et al.\ (\cite{Chen}) 
recently detected a faint silicate feature around 10~$\mu$m.

\paragraph{{\bf HD 123356:}}
\label{sect:hd123356}

Not much is known about \object{HD 123356}. In ``The Washington 
Visual Double Star Catalog'' (Worley \& Douglass \cite{Worley}) it is 
listed as having a companion 2.2~mag fainter in V-band, separated by 
2.5$\arcsec$ at a position angle of 126$\degr$. Mannings \& Barlow 
(\cite{Mannings98}) identified \object{HD 123356} as a candidate main 
sequence debris disk object, by cross-correlating the ``Michigan 
Catalog of Two-dimensional Spectral Types for the HD Stars'' with the 
``IRAS Faint Source Survey Catalog''. Sylvester et  
al.\ (\cite{Sylvester00}) show a moderate-SNR N-band spectrum which 
looks quite different from ours. The authors, however, clarify that 
their CGS3 beam of 5.5$\arcsec$ was pointing at the position of the 
primary (putting the secondary at the edge of the beam) and that not all 
the IR signal is detected if a substantial part of the emission arises
from the secondary. No significant CO emission (Dent et al.\ \cite{Dent}) 
and only a very low degree of NIR polarimetric emission (Hales et 
al.\ \cite{Hales}) was found. 

Compared to the other emission feature sources, the TIMMI2 N-band spectrum 
of \object{HD 123356} appears atypical, since the flux remains rather 
constant between 8 and 11~$\mu$m. However, our N-band spectra were 
observed on three occasions (March 2003, February 2005 and January 2006), 
always showing the same spectral shape. From modelling the circumstellar 
emission, 
the dust seems to be at a rather early processing stage (or mainly 
unprocessed). As shown in 
Fig.\,\ref{fig:decompo_ttauri}, the emission is dominated by small amorphous 
silicates, in addition to small amounts of forsterite, silica, enstatite 
and PAH.

While the currently available data of \object{HD 123356} are consistent
with a Vega-type star and circumstellar dust in a retarded stage of 
processing, the
real nature of this system can only be revealed with high-resolution
NIR- and MIR-observations.

\paragraph{{\bf HD 181296:}}

Mannings \& Barlow (\cite{Mannings98}) identified \mbox{\object{HD 181296}}
(alias \mbox{HR 7329}, or $\eta$~Tel) as a Vega-like
star. A brown dwarf companion at a projected distance of 200~AU was 
found by Lowrance et al.\ (\cite{Lowrance}). Also this target is part of 
the $\beta$~Pic moving group (Torres et al.\ \cite{Torres06}).

From our TIMMI2 N1 photometry at central wavelength 8.6~$\mu$m, the 
reported 12~$\mu$m IRAS flux appears overestimated, while our photometry 
and N-band spectral profile are in excellent agreement with the fluxes 
Mamajek et al.\ (\cite{Mamajek}) reported for 10.3~$\mu$m and 11.6~$\mu$m.

\subsection{Non-detections}

Two objects for which a significant MIR flux was reported by IRAS remain
undetected in our observations, although the stated fluxes should have been 
easily observable. Most likely this was caused by other sources in the 
large IRAS beam. We report these non-detections to clarify the situation 
for these targets in the literature.

\paragraph{{\bf HD 158153:}}

In the SIMBAD database an IRAS flux of 3.83~Jy at 12~$\mu$m is given.
This source was also mentioned in the work by 
Oudmaijer et al.\ (\cite{Oudmaijer}). With a 10\,$\sigma$ sensitivity of 
25~mJy/hour we did not detect a source at this position within 3 minutes of 
integration time, suggesting a flux limit $<$110~mJy. Several IRAS sources 
are located nearby, of which the 11$\arcsec$ 
close object \object{IRAS 17251-2320} seems to emit the largest flux. We 
find a rising dust continuum for that source, but do not present it here
in detail, as its nature (stellar or not) remains unclear.

\paragraph{{\bf SAO 185668}} 

is listed in SIMBAD with an IRAS flux of 1.61~Jy 
at 12~$\mu$m. Malfait et al.\ (\cite{Malfait98}) considered it a Vega-like 
star based on its IRAS photometry. With a poor 10\,$\sigma$ MIR sensitivity 
around 100--150~mJy/hour no signal was found within 12 minutes on-source 
time, resulting in a flux limit about $<$0.3~Jy.

\end{appendix}


\begin{thebibliography}{}

\bibitem[2004]{Acke04}
Acke, B. \& van den Ancker, M. E. 2004b, A\&A, 426, 151

\bibitem[2005]{Acke05}
Acke, B., van den Ancker, M. E., \& Dullemond, C. P. 2005, A\&A, 436, 209

\bibitem[2005]{Beichman}
Beichman, C. A., Bryden, G., Gautier, T. N., et al.\ 2005, ApJ, 626, 1061

\bibitem[2008]{Boersma}
Boersma, C., Bouwman, J., Lahuis, F., et al.\ 2008, A\&A, 484, 241

\bibitem[2001]{Bouwman01}
Bouwman, J., Meeus, G., de Koter, A., et al.\ 2001, A\&A, 375, 950

\bibitem[2008]{Bouwman08}
Bouwman, J., Henning, Th., Hillenbrand, L. A., et al.\ 2008, ApJ, 683, 479

\bibitem[1986]{Byrne}
Byrne, P. B. 1986, IrAJ, 17, 294

\bibitem[2007]{Carmona}
Carmona, A., van den Ancker, M. E., \& Henning, Th. 2007, A\&A, 464, 687

\bibitem[2007]{Catala}
Catala, C., Alecian, E., Donati, J.-F., et al.\ 2007, A\&A, 462, 293

\bibitem[2006]{Chen}
Chen, C. H., Sargent, B. A, Bohac, C., et al.\ 2006, ApJ Supp., 166, 351

\bibitem[1998]{Cohen}
Cohen, M. 1998, AJ, 115, 2092

\bibitem[1998]{Coulson}
Coulson, I. M., Walther, D. M., \& Dent, W. R. F. 1998, MNRAS, 296, 934

\bibitem[2000]{Angelo}
D'Angelo, G., Errico, L., Gomez, M. T., et al.\ 2000, A\&A, 356, 888

\bibitem[2005]{Dent}
Dent, W. R. F., Greaves, J. S., \& Coulson, I. M. 2005, MNRAS, 359, 663

\bibitem[2002]{Dommanget}
Dommanget, J., \& Nys, O. 2002, yCat, I/274

\bibitem[1995]{Dorschner}
Dorschner, J., Begemann, B., Henning, Th., Jaeger, C., \& Mutschke, H.
1995, A\&A, 300, 503

\bibitem[2007]{Doucet}
Doucet, C., Habart, E., Pantin, E., et al.\ 2007, A\&A, 470, 625

\bibitem[2004]{Dullemond}
Dullemond, C. P., \& Dominik, C. 2004, A\&A, 417, 159

\bibitem[1997]{Dunkin}
Dunkin, S. K., Barlow, M. J., Ryan, S. G. 1997, MNRAS, 290, 165

\bibitem[2007]{Folsom}
Folsom, C. P. 2007, ``Magnetic, chemical and rotational properties of 
the Herbig Ae/Be binary HD~72106'', Thesis at Queen's University; 
Kingston, Ontario, Canada

\bibitem[2006]{Furlan}
Furlan, E., Hartmann, L., Calvet, N., et al.\ 2006, ApJ Supp., 165, 568

\bibitem[2005]{Geers05}
Geers, V. C., Augereau, J.-C., Pontoppidan, K. M., et al.\ 2005,
in ``High Resolution Infrared Spec\-troscopy in Astronomy'';
ed. K\"aufl, H. U., Siebenmorgen, R., \& Moorwood, A. (Springer),
p.\ 239

\bibitem[2006]{Geers06}
Geers, V. C., Augereau, J.-C., Pontoppidan, K. M., et al.\ 2006, A\&A, 
459, 545

\bibitem[2007a]{Geers07a}
Geers, V. C., Pontoppidan, K. M., van Dishoeck, E. F., et al.\ 2007a, A\&A, 
469, L35

\bibitem[2007b]{Geers07b}
Geers, V. C., van Dishoeck, E. F., Visser, R., et al.\ 2007b, A\&A, 476, 279

\bibitem[2006]{Green}
Green, J. D., Hartmann, L., Calvet, N., et al.\ 2006, ApJ, 648, 1099

\bibitem[2004]{Habart04}
Habart, E., Testi, L., Natta, A., \& Carbillet, M. 2004, ApJ, 614, L129

\bibitem[2006]{Habart06}
Habart, E., Natta, A., Testi, L., \& Carbillet, M. 2006, A\&A, 449, 1067

\bibitem[2006]{Hales}
Hales, A. S., Gledhill, T. M., Barlow, M. J., \& Lowe, K. T. E. 2006,
MNRAS, 365, 1348

\bibitem[1997]{Hallenbeck}
Hallenbeck, S., \& Nuth, J. 1997, Ap\&SS, 255, 427

\bibitem[1998]{Hanner}
Hanner, M. S., Brooke, T. Y., \& Tokunaga, A. T. 1998, ApJ, 502, 871

\bibitem[1996]{Hartmann}
Hartmann, L., \& Kenyon, S. J. 1996, ARA\&A, 34, 207

\bibitem[1994]{Henning}
Henning, Th., Launhardt, R., Steinacker, J., \& Thamm, E. 1994, A\&A, 
291, 546

\bibitem[2005]{Herbig05}
Herbig, G. H. 2005, AJ, 130, 815

\bibitem[2004]{Hernandez}
Hern\'andez, J., Calvet, N., Brice\~no, C., Hartmann, L., \& Berlind, P. 
2004, AJ, 127, 1682

\bibitem[2004]{Higdon}
Higdon, S. J. U., Devost, D., Higdon, J. L., et al.\ 2004, PASP, 116, 975

\bibitem[1992]{Hillenbrand}
Hillenbrand, L. A., Strom, S. E., Vrba, F. J., \& Keene, J. 1992, ApJ, 
397, 613

\bibitem[2006]{Honda}
Honda, M., Kataza, H., Okamoto, Y. K., et al.\ 2006, ApJ, 646, 1024

\bibitem[2001]{Hony}
Hony, S., Van Kerckhoven, C., Peeters, E., et al.\ 2001, A\&A, 370,
1030

\bibitem[2004]{Houck}
Houck, J. R., Roellig, T. L., van Cleve, J., et al.\ 2004, ApJ Supp., 
154, 18

\bibitem[1999]{Hudgins}
Hudgins, D. M., \& Allamandola, L. J. 1999, ApJ, 516, L41

\bibitem[1990]{Hutchinson}
Hutchinson, M. G., Evans, A., Winkler, H., \& Spencer Jones, J. 1990, 
A\&A, 234, 230

\bibitem[1998]{Jaeger}
Jaeger, C., Molster, F.J., Dorschner, J., et al.\ 1998, A\&A, 339, 904

\bibitem[2001]{Jayawardhana}
Jayawardhana, R., Fisher, R. S., Telesco, C. M., et al.\ 2001, AJ, 122, 
2047

\bibitem[1997]{Jensen}
Jensen, E. L. N., \& Mathieu, R. D. 1997, AJ, 114, 301

\bibitem[2009]{Juhasz}
Juh\'asz, A., Henning, Th., Bouwman, J., et al.\ 2009, ApJ, in press, 
arXiv0902.0405

\bibitem[2003]{Kaufl}
K\"aufl, H.- U., Sterzik, M. F., Siebenmorgen, R., et al.\ 2003, SPIE, 
4841, 117

\bibitem[2004]{Kemper}
Kemper, F., Vriend, W.J., \& Tielens, A. G. G. M. 2004, ApJ, 609, 826

\bibitem[1988]{Kenyon88}
Kenyon, S. J., Hartmann, L., \& Hewett, R. 1988, ApJ, 325, 231

\bibitem[2000]{Kenyon00}
Kenyon, S. J., Kolotilov, E. A., Ibragimov, M. A., \& Mattei, J. A. 2000,
ApJ, 531, 1028

\bibitem[2006]{Kessler}
Kessler-Silacci, J. E., Augereau, J.-C., Dullemond, C. P., et al.\ 2006, 
ApJ, 639, 275

\bibitem[2007]{Kessler07}
Kessler-Silacci, J. E., Dullemond, C. P., Augereau, J.-C., et al.\ 2007, 
ApJ, 659, 680

\bibitem[1994]{Kurucz}
Kurucz, R. L. 1994, ``Solar abundance model atmospheres for 0, 1, 2, 4, 8
km/s'', CD-ROM No.\ 19. (Cambridge, Mass.: Smith\-sonian Astro\-physical 
Observatory)

\bibitem[2004]{Lagage}
Lagage, P. O., Pel, J. W., Authier, M., et al.\ 2004, Msngr, 117, 12

\bibitem[2006]{Lagage06}
Lagage, P. O., Doucet, C., Pantin, E., et al.\ 2006, Science, 314, 621

\bibitem[1997]{Leinert97}
Leinert, C., Richichi, A., \& Haas, M. 1997, A\&A, 318, 472

\bibitem[2004]{Leinert04}
Leinert, Ch., van Boekel, R., Waters, L. B. F. M., et al.\ 2004, A\&A, 
423, 537

\bibitem[2002]{Li}
Li, A., \& Draine, B. T. 2002, ApJ, 572, 232

\bibitem[2008]{Lisse}
Lisse, C. M., Chen, C. H., Wyatt, M. C., \& Morlok, A. 2008, ApJ, 673, 1106

\bibitem[2006]{Lovis}
Lovis, C., Mayor, M., Pepe, F., et al.\ 2006, Nature, 441, 305

\bibitem[2000]{Lowrance}
Lowrance, P. J., Schneider, G., Kirkpatrick, J. D., et al.\ 2000, ApJ, 
541, 390

\bibitem[1998]{Malfait98}
Malfait, K., Bogaert, E., \& Waelkens, C. 1998, A\&A, 331, 211

\bibitem[2004]{Mamajek}
Mamajek, E. E., Meyer, M. R., Hinz, P. M., et al.\ 2004, ApJ, 612, 496

\bibitem[1998]{Mannings98}
Mannings, V., \& Barlow, M. J. 1998, ApJ, 497, 330

\bibitem[2005]{Mattioda}
Mattioda, A. L., Hudgins, D. M., \& Allamandola, L. J. 2005, ApJ, 629, 
1188

\bibitem[2001]{Meeus01}
Meeus, G., Waters, L. B. F. M., Bouwman, J., et al.\ 2001, A\&A, 365, 476

\bibitem[2002]{Meeus02}
Meeus, G., Bouwman, J., Dominik, C., Waters, L.  B.  F.  M., \& de 
Koter, A. 2002, A\&A, 392, 1039 

\bibitem[2003]{Meeus03}
Meeus, G., Sterzik, M., Bouwman, J., \& Natta, A. 2003, A\&A, 409,~L25

\bibitem[2007]{Meyer}
Meyer, M. R., Backman, D. E., Weinberger, A. J., \& Wyatt, M. C. 2007,
in ``Protostars and Planets V''; ed. B. Reipurth, D. Jewitt, \& K. Keil 
(University of Arizona Press, Tucson), p.\ 573

\bibitem[1998]{Moorwood}
Moorwood, A., Cuby, J. G., \& Lidman, C. 1998, Msngr, 91, 9

\bibitem[2004]{Natta}
Natta, A., Testi, L., Neri, R., Shepherd, D. S., Wilner, D. J. 2004, A\&A,
416, 179

\bibitem[2004]{Okamoto}
Okamoto, Y. K., Kataza, H., Honda, M., et al.\ 2004, Nature, 431, 660

\bibitem[1992]{Oudmaijer}
Oudmaijer, R. D., van der Veen, W. E. C. J., Waters, L. B. F. M., et 
al.\ 1992, A\&A Supp., 96, 625

\bibitem[2008]{Pascucci}
Pascucci, I., Apai, D., Hardegree-Ullman, E. E., et al.\ 2008, ApJ, 673, 477

\bibitem[1991]{Pasquini}
Pasquini, L., Cutispoto, G., Gratton, R., \& Mayor, M. 1991, A\&A, 248, 72

\bibitem[2002]{Peeters02}
Peeters, E., Hony, S., Van Kerckhoven, C., et al.\ 2002, A\&A, 390, 1089

\bibitem[2004a]{Peeters04a}
Peeters, E., Allamandola, L.~J., Hudgins, D.~M., Hony, S., \& Tielens, 
A.~G.~G.~M. 2004a, in Astronomical Society of the Pacific Conference Series, 
Vol.\ 309, Astrophysics of Dust, ed. A.~N. Witt, G.~C. Clayton, \& B.~T. 
Draine, p.\ 141

\bibitem[2004b]{Peeters04}
Peeters, E., Mattioda, A. L., Hudgins, D. M., \& Allamandola, L. J. 2004b,
ApJ, 617, L65 

\bibitem[2006]{Perrin}
Perrin, M. D., Duchene, G., Kalas, P., \& Graham, J.R. 2006, ApJ,  645, 1272

\bibitem[2005]{Pogodin} 
Pogodin, M. A., Franco, G. A. P., \& Lopes, D. F. 2005, A\&A, 438, 239

\bibitem[2001]{Prato}
Prato, L., Ghez, A. M., Pi\~na, R. K., et al.\ 2001, ApJ, 549, 590

\bibitem[2003]{Przygodda}
Przygodda, F., van Boekel, R., \`Abrah\`am, P., et al.\ 2003, A\&A, 412,
L43

\bibitem[2006]{Quanz}
Quanz, S. P., Henning, Th., Bouwman, J., Ratzka, Th., \& Leinert, Ch. 
2006, ApJ, 648, 472

\bibitem[2007]{Quanz07}
Quanz, S. P., Henning, Th., Bouwman, J., et al.\ 2007, ApJ, 668, 359

\bibitem[2000]{Quast}
Quast, G. R., Torres, C. A. O., de La Reza, R., da Silva, L., \& Mayor, 
M. 2000, IAUS 200, 28

\bibitem[1989]{Rietmeijer}
Rietmeijer, F. J. M. 1989, in ``Lunar and Planetary Science Con\-ference, 
19th'', Proceedings A89-36486 15-91 (Cambridge University Press / 
Lunar and Planetary Institute), p.\ 513

\bibitem[1979]{Savage}
Savage, B. D., Mathis, J. S. 1979, ARA\&A, 17, 73 

\bibitem[2006]{Schegerer}
Schegerer, A., Wolf, S., Voshchinnikov, N. V., Przygodda, F., \&
Kessler-Silacci, J. E. 2006, A\&A, 456, 535

\bibitem[2005a]{paper1}
Sch\"utz, O., Meeus, G., \& Sterzik, M. F. 2005a, A\&A, 431, 165

\bibitem[2005b]{paper2}
Sch\"utz, O., Meeus, G., \& Sterzik, M. F. 2005b, A\&A, 431, 175

\bibitem[2005c]{Schuetz}
Sch\"utz, O., \& Sterzik, M. F. 2005c, 
in ``High Resolution Infrared Spec\-troscopy in Astronomy'';
ed. K\"aufl, H. U., Siebenmorgen, R., \& Moorwood, A. (Springer),
p.\ 104

\bibitem[1973]{Servoin}
Servoin, J.L., \& Piriou, B. 1973, Phys. Stat. Sol. B, 55, 677

\bibitem[2004]{Sheret}
Sheret, I., Dent, W. R. F., \& Wyatt, M. C. 2004, MNRAS, 348, 1282

\bibitem[2007]{Sicilia}
Sicilia-Aguilar, A., Hartmann, L. W., Watson, D., et al.\ 2007,
ApJ, 659, 1637

\bibitem[2005]{Sloan}
Sloan, G. C., Keller, L. D., Forrest, W. J., et al.\ 2005, ApJ, 632, 956

\newpage

\bibitem[2007]{Sloan07}
Sloan, G. C., Jura, M., Duley, W. W., et al.\ 2007, ApJ, 664, 1144

\bibitem[2004]{Smith}
Smith, T. L., Clayton, G. C., \& Valencic, L. 2004, AJ, 128, 357

\bibitem[2001]{Song01}
Song, I., Caillault, J.-P., Barrado y Navascu\'es, D., \& Stauffer,
J. R. 2001, ApJ, 546, 352

\bibitem[2005]{Song05}
Song, I., Zuckerman, B., Weinberger, A. J., \& Becklin, E. E. 2005, Nature, 
436, 363

\bibitem[1961]{Spitzer}
Spitzer, W.G., \& Kleinman, D.A. 1961, Phys. Rev., 121, 1324

\bibitem[1991]{Stencel}
Stencel, R. E. \& Backman, D. E. 1991, ApJ Supp., 75, 905

\bibitem[2005]{Sterzik}
Sterzik, M. F., Melo, C. H. F., Tokovinin, A. A., \& van der Bliek, N. 
2005, A\&A, 434, 671

\bibitem[2000]{Strassmeier}
Strassmeier, K. G., \& Rice, J. B. 2000, A\&A, 360, 1019

\bibitem[1996]{Sylvester96}
Sylvester, R. J., Skinner, C. J., Barlow, M. J., \& Mannings, V. 1996,
MNRAS, 279, 915

\bibitem[2000]{Sylvester00}
Sylvester, R. J., \& Mannings, V. 2000, MNRAS, 313, 73

\bibitem[2001]{Sylvester01}
Sylvester, R. J., Dunkin, S. K., \& Barlow, M. J. 2001, MNRAS, 327,~133

\bibitem[2008]{Tielens}
Tielens, A. G. G. M. 2008, to appear in Annual Review of Astronomy and 
Astrophysics, in press

\bibitem[1995]{Torres95}
Torres, C. A. O., Quast, G., de la Reza, R., Gregorio-Hetem, J. \& 
L\'epine, J. R. D. 1995, AJ, 109, 2146

\bibitem[2004]{Torres04}
Torres, G. 2004, AJ, 127, 1187

\bibitem[2006]{Torres06}
Torres, C. A. O., Quast, G. R., da Silva, L., et al.\ 2006, A\&A, 460, 695

\bibitem[2003]{Boekel03}
van Boekel, R., Waters, L. B. F. M., Dominik, C., et al.\ 2003, A\&A, 400, 
L21

\bibitem[2004]{Boekel04}
van Boekel, R., Min, M., Leinert, Ch., et al.\ 2004, Nature, 432, 479

\bibitem[2005]{Boekel05}
van Boekel, R., Min, M., Waters, L. B. F. M., et al.\ 2005, A\&A, 437, 189

\bibitem[1998]{Ancker}
van den Ancker, M. E., de Winter, D., \& Tjin A Djie, H. R. E. 1998, A\&A, 
330, 145

\bibitem[2004]{Diedenhoven}
van Diedenhoven, B., Peeters, E., Van Kerckhoven, C., et al.\ 2004, ApJ, 
611, 928 

\bibitem[2000]{Kerckhoven}
Van Kerckhoven, C., Hony, S., Peeters, E., et al.\ 2000, A\&A, 357, 1013

\bibitem[2002]{Kerckhoven02}
Van Kerckhoven, C. 2002, PhD thesis at Catholic University of Leuven, 
Belgium

\bibitem[2003]{Vieira}
Vieira, S.L.A., Corradi, W.J.B., Alencar, S.H.P., et al.\ 2003, AJ,
126, 2971

\bibitem[2004]{Wang}
Wang, H., Apai, D., Henning, T., \& Pascucci, I. 2004, ApJ, 601, L83

\bibitem[2004]{Werner}
Werner, M. W., Roellig, T. L., Low, F. J., et al.\ 2004, ApJ Supp., 154, 1

\bibitem[1997]{Worley}
Worley C. E., \& Douglass G. G. 1997, A\&A Supp., 125, 523

\bibitem[2005]{Wyatt}
Wyatt, M. C., Greaves, J. S., Dent, W. R. F., \& Coulson, I. M. 2005, ApJ, 
620, 492

\bibitem[2000]{Yudin}
Yudin, R. V. 2000, A\&A Supp., 144, 285

\bibitem[2001]{Zuckerman01}
Zuckerman, B., Song, I., Bessell, M. S., \& Webb, R. A. 2001, ApJ, 562, L87

\bibitem[2004]{Zuckerman04}
Zuckerman, B., \& Song, I. 2004, ARA\&A, 42, 685

\end{thebibliography}
\end{document}